%% file: paper-mars-2020.tex
\documentclass[submission,copyright,creativecommons]{eptcs}
\providecommand{\event}{MARS 2020} 
\input{convecs-3}

\input{macr_pic}

\usepackage{listings}
\usepackage{enumerate}
\usepackage{xspace}
\usepackage{soul} % to highlight text
\usepackage{multirow}
\usepackage{epsfig,color}

\definecolor{mygreen}{rgb}{0,0.6,0}
\definecolor{mygray}{rgb}{0.5,0.5,0.5}
\definecolor{mymauve}{rgb}{0.1,0.2,0.7}
\definecolor{olivegreen}{cmyk}{.6,.4,0.8,0}

\lstdefinelanguage{Lustre}
  {morekeywords={and,as,begin,constraint,do,done,else,end,exist,fby,for,if,in,let,loop,node,not,of,or,pre,returns,select,tel,then,to,when,where,while,with,function,type,enum,struct,const,var,restart,SSM,:,set,check,uncheck,cycle},
   keywordstyle=\color{mymauve}\bfseries,
    morecomment=[n]{(*}{*)},
    basicstyle=\small\ttfamily,
    morecomment=[l]{--},
    commentstyle=\rmfamily\itshape\color{olivegreen}
  }[keywords,comments]
 
\makeatletter
\def\mylabel#1#2{\@bsphack\if@filesw {\let\thepage\relax
   \def\protect{\noexpand\noexpand\noexpand}%
   \edef\@tempa{\write\@auxout{\string
      \newlabel{#1}{{#2}{\thepage}{}{figure.1.1}{}}}}%
   \expandafter}\@tempa
   \if@nobreak \ifvmode\nobreak\fi\fi\fi\@esphack}
\makeatother

\title{Specifying a Cryptographical Protocol in Lustre and SCADE}

\author{
Lina Marsso
\null \\
Univ. Grenoble Alpes, INRIA, CNRS, LIG, Grenoble, France \\
E-mail: lina.marsso@inria.fr
}

\def\titlerunning{Specifying a Cryptographical Protocol in Lustre and SCADE}
\def\authorrunning{L. Marsso}
\begin{document}
\sloppy
\maketitle

\begin{abstract}
We present SCADE and Lustre models of the Message Authenticator Algorithm (MAA), which is one of the first cryptographic functions for computing a message authentication code.
The MAA was adopted between 1987 and 2001, in international standards (ISO 8730 and ISO 8731-2), to ensure the authenticity and integrity of banking transactions.
This paper discusses the choices and the challenges of our MAA implementations.
Our SCADE and Lustre models validate 201 official test vectors for the MAA.
\end{abstract}

\section{Introduction}
%The MAA
One of the first Message Authentication Code (MAC) algorithms to gain widespread acceptance was the Message Authenticator Algorithm (MAA, also known as the Message Authentication Algorithm)~\cite{Davies-85,Davies-Clayden-88,Preneel-11} designed in 1983 by Donald Davies and David Clayden. 
The MAA was adopted by ISO in 1987 and became part of the international standards 8730~\cite{ISO-8730:1986} and 8731-2~\cite{ISO-8731-1:1987}. 
Later, cryptanalysis of the MAA revealed various weaknesses, including feasible brute-force attacks, existence of collision clusters, and key-recovery techniques~\cite{Preneel-vanOorschot-96,Rijmen-Preneel-DeWin-96,Preneel-Rumen-vanOorschot-97,Preneel-vanOorschot-99,Preneel-11}. 
For this reason, the MAA was withdrawn from ISO standards in 2002.
Even if the MAA was deprecated, as it happens to every cryptographic protocol as time passes, various formal specifications of the MAA have been developed before and after 2002.
Most likely because the MAA wqs the first message authentication code algorithm and its definition is stable and will not change anymore.
Moreover, it was designed as a standalone algorithm that does not rely on any preexisting hash function or cipher and, hence, has manageable complexity.
The existing formal specifications of the MAA are either non-executable, such as the VDM one in 1990~\cite{Parkin-ONeill-90,Parkin-ONeill-91}, the Z one in 1991~\cite{Lai-91}, the LOTOS one in 1990/1991~\cite{Munster-91-a,Munster-91-b} or executable, such as the LOTOS one (\cite{Garavel-Marsso-18}), the LNT one (\cite{Garavel-Marsso-18}), and those automatically derived from a term rewrite system (\cite{Garavel-Marsso-17}).
More details about these earlier formal specifications of the MAA are available in~\cite{Garavel-Marsso-18}.
Thus, the MAA could be viewed as a sort of Rosetta stone\footnote{The Rosetta Code repository (\url{http://www.rosettacode.org/wiki/Rosetta_Code}) collects classical algorithm implementations, written in hundreds different programming languages.} for formal methods. However, synchronous models of the MAA have been missing
so far, and this is what the present article is about.

% contribution
This paper brings MAA models in synchronous languages to enrich the collection of formal models for the MAA algorithm.
More precisely, in this paper, we present a formal model of the MAA with the synchronous formal language Lustre~\cite{Halbwachs-Caspi-Raymond-Pilaud-91} dedicated to design embedded control systems; and with the industrial formal language SCADE (Safety Critical Application Development Environment)~\cite{Colaco-Pagano-Pouzet-17}.
We discuss implementation choices and challenges, common to both models, partly because the SCADE language is based on Lustre language.
Finally, we explain how we test the two formal models with the complete set of test vectors derived from the initial specification~\cite{Davies-Clayden-88}.
% Plan of the paper
The remainder of the paper is organized as follows. 
Section~\ref{ch:MAA} presents the technical perspective of the MAA. 
The modeling of the MAA using the synchronous language SCADE/Lustre is described in Section~\ref{sec:lustre}.  
Section~\ref{sec:validation} precises how the SCADE and the Lustre models have been validated. 
Finally, Section~\ref{sec:conc} gives some concluding remarks and future work directions.

% ----------------------------------------------------------------------
% Explanantion of the MAA algorithm.
\section{The MAA Algorithm}
\label{ch:MAA}
% intro of the section
In data security, a Message Authentication Code (MAC) is a short sequence of bits that is computed from a given message; a MAC ensures both the authenticity and integrity of the message, i.e., that the message sender is the stated one and that the message contents have not been altered.
The design of MAC usually involves cryptographic keys shared between the message sender and receiver, which makes it more secure under attacks.
In this section, we briefly explain the principles of the MAA while more detailed explanations are provided in \cite{Davies-85,Davies-Clayden-88} and~\cite[Algorithm~9.68]{Menezes-vanOorschot-Vanstone-96}.

The MAA was intended to be implemented in software and to run on 32-bit computers. 
Hence, its design intensively relies on 32-bit words (called blocks) and 32-bit machine operations.
The MAA takes as inputs a key and a message. 
The key has 64 bits and is split into two blocks \emph{J} and \emph{K}.
The message is seen as a sequence of blocks.
If the number of bytes of the message is not a multiple of four, extra null bytes are added at the end of the message to complete the last block.
The size of the message should be less than 1,000,000 blocks; otherwise, the MAA result is said to be undefined; we believe that this restriction, which is not inherent to the algorithm itself, was added in the second ISO standard~\cite{ISO-8731-2:1992} to provide MAA implementations with an upper bound (four megabytes) on the size of memory buffers used to store messages.

The MAA produces as output a block, which is the MAC value computed from the key and the message. 
The fact that this result has only 32 bits proved to be a major weakness enabling cryptographic attacks; MAC values computed by modern algorithms now have a much larger number of bits. 
Apart from the aforementioned restriction on the size of messages, the MAA behaves as a totally-defined function; its result is deterministic in the sense that, given a key and a message, there is only a single MAC result, which neither depends on implementation choices nor on hidden inputs, such as randomly-generated numbers.

The MAA calculations rely upon conventional 32-bit logical and arithmetic operations, among which: \lstinline+AND+ (conjunction), \lstinline+OR+ (disjunction), \lstinline+XOR+ (exclusive disjunction), \lstinline+CYC+ (circular rotation by one bit to the left), \lstinline+ADD+ (addition), \lstinline+CAR+ (carry bit generated by 32-bit addition), \lstinline+MUL+ (multiplication, sometimes decomposed into \lstinline+HIGH_MUL+ and \lstinline+LOW_MUL+, which denote the most- and least-significant blocks in the 64-bit product of a 32-bit multiplication).
On this basis, more involved operations are defined, among which \lstinline+MUL1+ (result of a 32-bit multiplication modulo $2^{32} - 1$), \lstinline+MUL2+ (result of a 32-bit multiplication modulo $2^{32} - 2$), \lstinline+MUL2A+ (faster version of \lstinline+MUL2+), \lstinline+FIX1+ and \lstinline+FIX2+ (two unary functions\footnote{The names \lstinline+FIX1+ and \lstinline+FIX2+ appeared in~\cite[pages~36 and~77]{Munster-91-a}.} respectively defined as $x \to$ \lstinline+AND+(\lstinline+OR+($x$,A),C) and $x \to$ \lstinline+AND+(\lstinline+OR+($x$,$B$),$D$), where $A$, $B$, $C$, and $D$ are the four hexadecimal block constants $A$ = \lstinline+02040801+, $B$ = \lstinline+00804021+, $C$ = \lstinline+BFEF7FDF+, and $D$ = \lstinline+7DFEFBFF+).
The MAA operates in three successive phases:
\begin{itemize}
  \item The \lstinline+PRELUDE+ takes the two blocks $J$ and $K$ of the key and converts them into six blocks $X_0$, $Y_0$, $V_0$, $W$, $S$, and $T$. 
  This phase is executed once. 
  After the prelude, $J$ and $K$ are no longer used.
  \item The \lstinline+MAIN_LOOP+ successively iterates on each block $M_n$ of the message ($M_1$, ..., $M_n$). This phase maintains three variables $X$, $Y$, and $V$ (initialized to $X_0$, $Y_0$, and $V_0$, respectively), which are modified at each iteration. 
The main loop also uses the value of $W$, but does not use $S$ and $T$.
  \item The \lstinline+CODA+ adds the blocks $S$ and $T$ at the end of the message and performs two more iterations on these blocks. 
  After the last iteration, the MAA result is XOR($X$, $Y$), called $Z$.
\end{itemize}

In 1987, the second ISO standard~\cite[Section~5]{ISO-8731-1:1987} introduced an additional feature (called mode of operation), which concerns messages longer than 256 blocks, which was not present in the early MAA versions designed at NPL.
\begin{figure}
\centering
%\Figure{maasegments}
\includegraphics[scale=0.3]{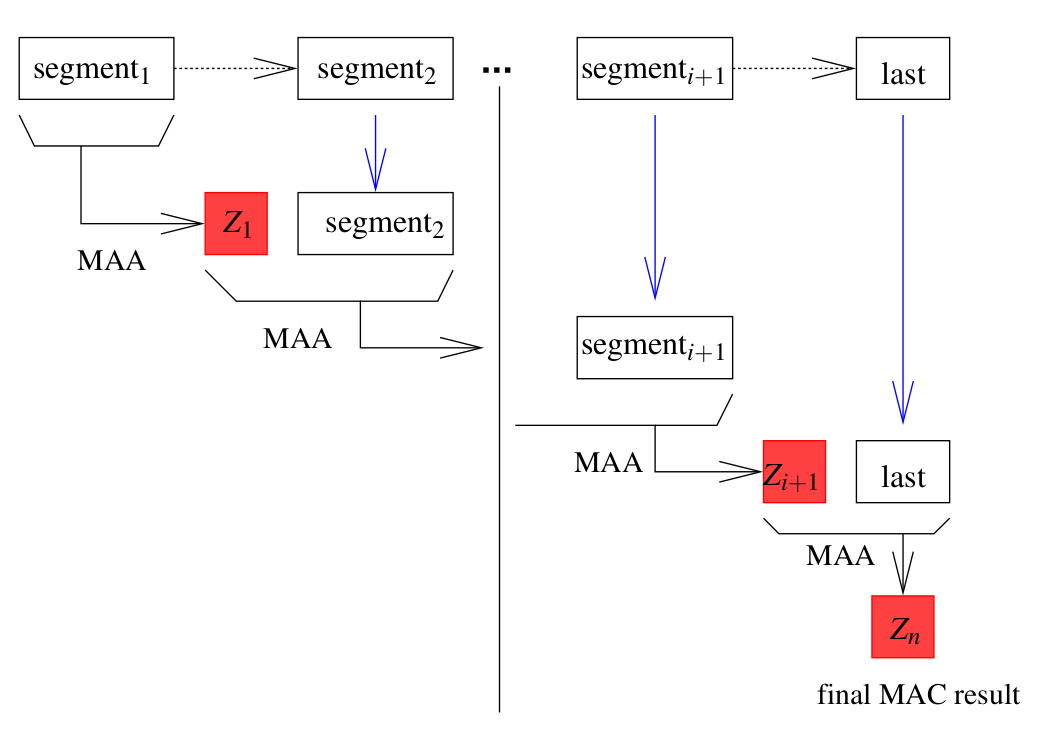}
\caption{Mode of operation of the MAA}
\label{fig:maaop}
\end{figure}
Figure~\ref{fig:maaop} gives an overview of ``the mode of operation''.
Each message longer than 256 blocks must be split into segments of 256 blocks each, with the last segment possibly containing less than 256 blocks.
The above MAA algorithm (prelude, main loop, and coda) is applied to the first segment, resulting in a value noted $Z_1$.
This block $Z_1$ is then inserted before the first block of the second segment, leading to a 257-block message to which the MAA algorithm is applied, resulting in a value noted $Z_2$. This is done repeatedly for all the $n$ segments, the MAA result $Z_i$ computed for the $i$-th segment being inserted before the first block of the $(i+1)$-th segment. 
Finally, the MAC for the entire message is the MAA result $Z_n$ computed for the last segment.
% add and MAC (which splits a message into 1024-byte segments and computes the overall signature of this message by iterating on each segment, the 4-byte signature of each segment being prepended to the bytes of the next segment)

% ----------------------------------------------------------------------
% presentation model Lustre

\section{Modelling MAA in SCADE and Lustre}
\label{sec:lustre}
% intro of the section
We consider the MAA models LNT-18~\cite{Garavel-Marsso-18} and REC-17~\cite{Garavel-Marsso-17} and extend them to model the full MAA algorithm in the synchronous  languages Lustre~\cite{Halbwachs-Caspi-Raymond-Pilaud-91} and SCADE~\cite{Berry-07}.
Lustre is a synchronous language based on the dataflow model of computation, which defines a reactive system that receives inputs at each tick of a logical clock and calculates outputs. 
The relation between inputs and outputs is described using a system of (unordered) equations. 
At a given cycle, the value of inputs can be memorized to be reused in the next cycles. 
The outputs at a given instant may thus depend on current and past inputs, but not on future inputs. 
The relations between inputs and outputs are specified in nodes using constants, operators, functions, auxiliary variables, as well as other nodes, which provide for a compositional description of complex systems.
SCADE is an industrial language that merges two different academic languages: Lustre, a declarative dataflow language, and Esterel~\cite{Berry-92} an imperative, synchronous process calculus; more precisely, SCADE embodies a graphical variant of Esterel known as SyncCharts~\cite{Andre-04}.
Although these languages are quite different, the designers of SCADE managed to integrate both in one coherent language. 
In the present article, our modelling is based upon the dataflow subset of SCADE, with little or no use of the Esteral/SynchCharts features.

Also, following the approach of a previous modelling of the MAA using term rewrite systems (\cite{Garavel-Marsso-17}), we chose not to rely on predefined data types of SCADE (bits, machine words, etc.) but to define all our types from scratch, using the type constructors provided by the language. 
One advantage of this approach is that it is fully formal, without implicit assumptions about the implementation of machine words or the behaviour of predefined
arithmetic operations. This also provides a thorough exercise to assess the correctness and performance of Lustre and SCADE compilers.

A large part of our SCADE/Lustre models contain generic definitions independent of the MAA that could be reused to model other examples. 
The whole SCADE model of the MAA is presented in both the Appendix~\ref{ap:maa} and the MARS repository, while the Lustre model will only appear in the MARS repository, as it is very similar to the SCADE one.
As in~\cite{Garavel-Marsso-17}, we choose to represent blocks as words of four bytes, rather than thirty-two bits.
Thereby, the logical operations on blocks (\lstinline+AND+, \lstinline+OR+, \lstinline+XOR+, and \lstinline+CYC+) are easy to define using bitwise and bytewise manipulations.
The bits, bytes and blocks are represented with the three following SCADE data structures:
\begin{lstlisting}[language=Lustre]
type Bit = enum {X0,X1};

type Octet = {x1: Bit; x2: Bit; x3: Bit; x4: Bit; 
              x5: Bit; x6: Bit; x7: Bit; x8: Bit};

type Block = {o1: Octet; o2: Octet; 
              o3: Octet; o4: Octet};
\end{lstlisting}
Next, we define a set of functions to implement the corresponding logical operations on bits, bytes, and blocks.
The arithmetical operations (\lstinline+ADD+, \lstinline+CAR+, and \lstinline+MUL+) have been implemented using 8-bit, 16-bit, and 32-bit adders and multipliers, more or less inspired from the theory of digital circuits.
Thus, the structure \lstinline+Pair+ shown below, represents the result of the multiplication of two blocks.
\begin{lstlisting}[language=Lustre]
type Pair = {w1: Block; w2: Block};
\end{lstlisting}
% The structures corresponding to the adders and multipliers
%\begin{lstlisting}[language=Lustre]
%type OctetSum = struct {x: Bit; o: Octet};
%type Half = struct {o1: Octet; o2: Octet};
%type HalfSum = struct {x: Bit; h: Half};
%type BlockSum = struct {x: Bit; w: Block};
%\end{lstlisting}
%TODO: READ BELLOW

\begin{figure}
\centering
\includegraphics[scale=0.3]{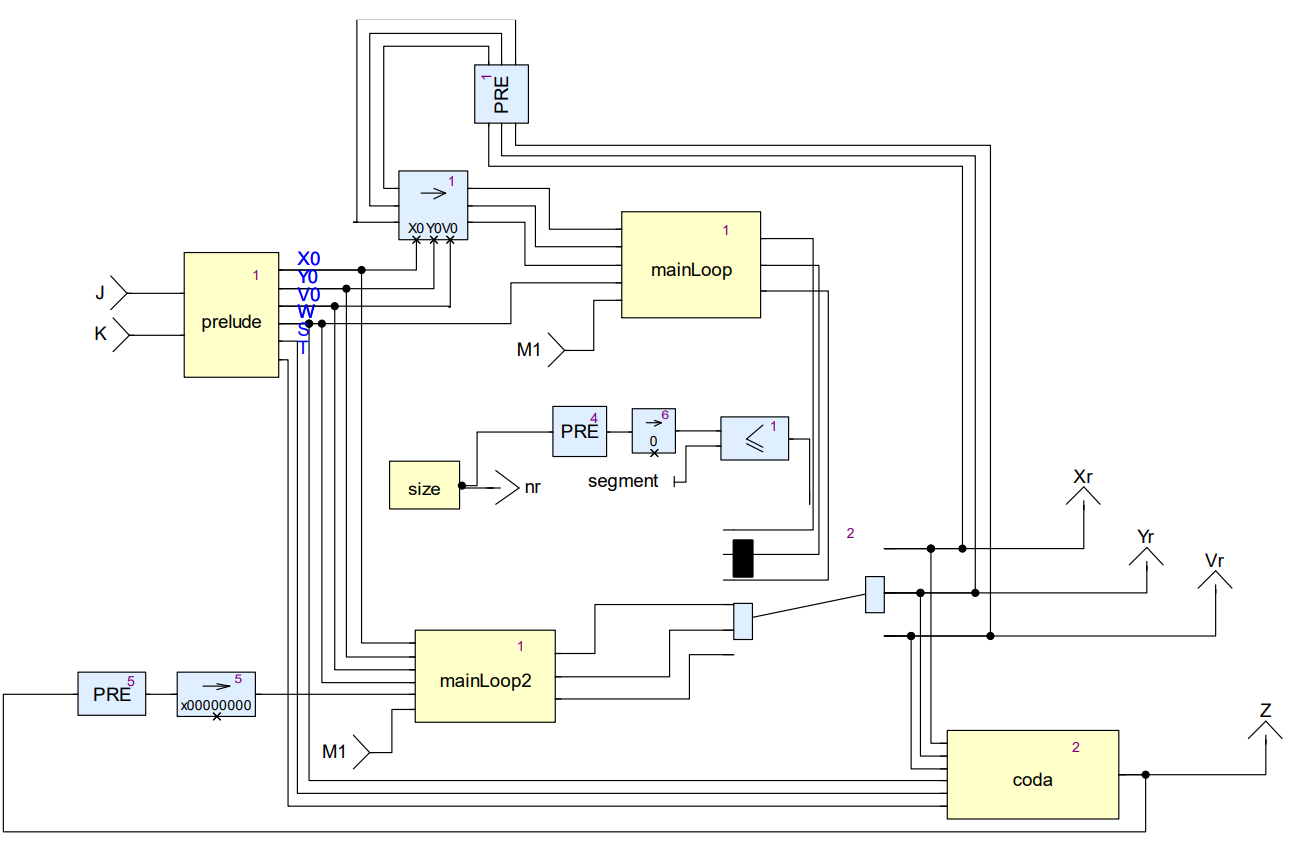}
\caption{Graphical SCADE node: MAC  with the ``mode of operation''.}
\label{fig:maac}
\end{figure}

A message is a list of blocks, each block of the message is an input of the main \lstinline+MAC+ node in SCADE, as illustrated in Figure~\ref{fig:maac} (graphically) and in Figure~\ref{fig:MAC} (in code). 
Note that in the synchronous dataflow SCADE/Lustre, a node describes the relation between its input and output parameters using a system of equations.
A node is composed of a set of unordered equations that define each output parameter according to the actual or previous input parameters.
%Long messages (i.e., containing more than one block) are hard coded in functions, as for instance, the twenty-words message in the Appendix~\ref{ap:20words}.

% MAA6
\begin{figure}[t]
\centering
\includegraphics[scale=0.58]{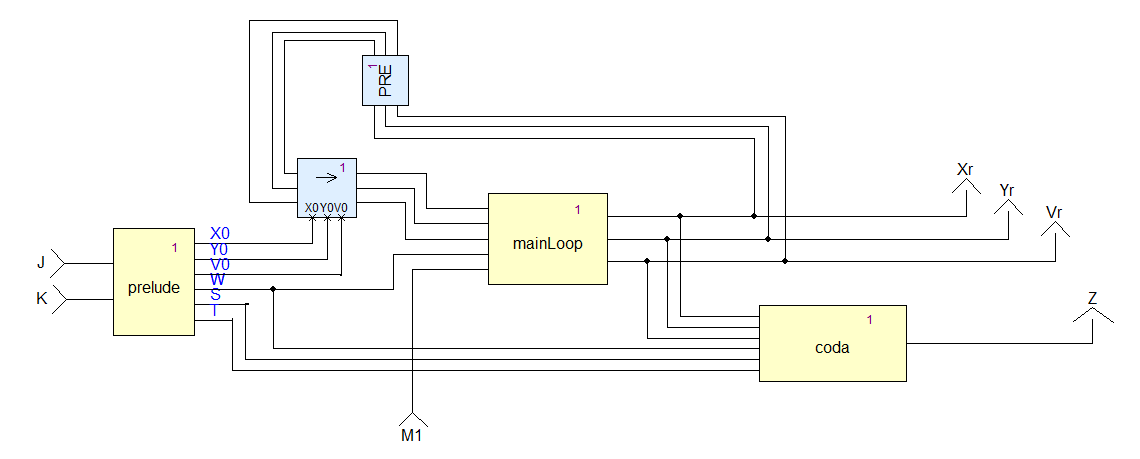}
\caption{Graphical SCADE node: MAC  without the ``mode of operation''.}
\label{fig:maasimple}
\end{figure}

The \lstinline+MAC+ node takes as input the message to encode, and a key, which corresponds to two words (called K and J).
A local variable \lstinline+init+ defines the beginning of a message, and another variable \lstinline+n+ stores the number of blocks, in order to ensure the implementation of the ``mode of operation'' of the MAA, i.e., segmentation of messages larger than 1024 bytes (as explained in Section~\ref{ch:MAA}).
The MAC provides as outputs the result of the intermediate computations (\lstinline+prelude+, \lstinline+mainloop+) in the auxiliary blocks (\lstinline+X, Y, V, W, S, T+), and the result of its main computation, i.e., the MAC for the given input (one key and a message), which is represented by a variable (\lstinline+Z+).
At each call of the SCADE node \lstinline+MAC+, one new block of the message is processed as follows:
\begin{itemize}
  \item for the $1^{st}$ block of the message, the auxiliary variables (\lstinline+X+, \lstinline+Y+, and \lstinline+V+) are computed with an iteration of the \lstinline+mainLoop+ function with the initial values (\lstinline+X0, Y0, V0+) computed by the function \lstinline+prelude+ shown in Figure~\ref{fig:MAC};
  \item for the $257^{th}$ block of the message, the auxiliary variables are computed with an iteration of the \lstinline+mainLoop2+ function, which consists of two iterations of the \lstinline+mainLoop+, one on the result of the previous coda (\lstinline+Z+), and one on this block message; and,
  \item for the other blocks of the message, the auxiliary variables are computed by taking into account the previous auxiliary variables (\lstinline+pre X+, \lstinline+pre Y+, and \lstinline+pre V+) computed by the previous block.
\end{itemize}
Figure~\ref{fig:maasimple} represents the graphical SCADE MAC\_A node without the implementation of the ``mode of operation'' (segmentation of the 1024 bytes messages).
User defined operators such as functions (\lstinline+prelude+, \lstinline+mainLoop+, and \lstinline+coda+) or nodes are represented by yellow boxes
while native SCADE operators are represented in blue boxes.
In particular, the two specific flow manipulation operators for nodes (inherited from Lustre) are used in the MAC\_A node (See Figure~\ref{fig:maasimple}):
\begin{itemize}
 \item the memory operator \lstinline[language=Lustre]+pre+ (``previous'') refers to the value of an input or output variable at the previous cycle;
 \item the initialization operator \lstinline[language=Lustre]+->+ (``followed by'') initializes a stream.
\end{itemize}
% MAA6
\begin{figure}[t]
\centering
\includegraphics[scale=0.2]{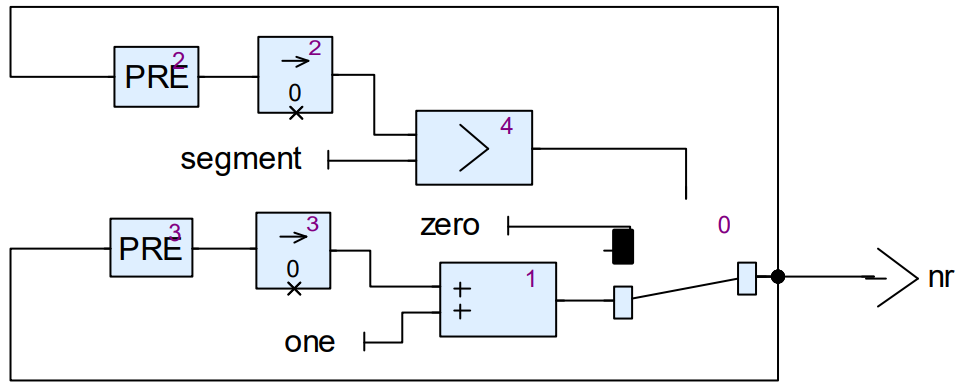}
\caption{The count node}
\label{fig:size}
\end{figure}
In order to implement the ``mode of operation'', we add a graphical node counting the number of blocks (\lstinline[language=Lustre]+nr+) of the segment, represented in Figure~\ref{fig:size}.
The node \lstinline+size+ contains the graphical choice operator, enabling the design of two branches.
The first one resets the number of blocks (\lstinline+nr+) for the $257^{th}$ block of the message (\lstinline+pre segment+), while the second branch increases \lstinline+nr+ for the other blocks.
Note that \lstinline+segment+ is a constant corresponding to the maximum size of a message segment (256 blocks).
The choice operator and the \lstinline+size+ node are then used to design the two branches of the graphical MAC node implementing the ``mode of operation'' represented in Figure~\ref{fig:maasimple}. 
Note that one of the branches consist of the MAC\_A node represented in Figure~\ref{fig:maasimple}.
We could use the modular reset feature of SCADE (\lstinline[language=Lustre]+restart+) to reinitialize the MAC node from the outside. 
It would have simplified the counter definition, and so the model, but since Lustre does not provide this feature, we don't use it for either of the models.
% MAA6
\begin{figure*}
  \makeatletter%
  \hspace*{\dimexpr 1em+\lst@numbersep}%
  \makeatother%
\begin{minipage}{.4\textwidth}
\begin{lstlisting}[language=Lustre]
node MAC (KJ: Key; Mn: Block; init: bool) 
         returns (X, Y, V, W, S, T, Z: Block; n: int32)
var X0, Y0, V0: Block; 
    newSegment: bool;
let
  n = 0 -> if init then 0 else ((pre n) + 1) mod 256;
  newSegment = false -> if pre n = 255 then true else false;
  -- initialisations
  X0, Y0, V0, W, S, T = prelude (KJ.J, KJ.K);
  -- mainloops
  X, Y, V = mainLoop (X0, Y0, V0, W, Mn) -> 
              if init then 
                mainLoop (X0, Y0, V0, W, Mn)
              else if newSegment then
                -- mode of operations
                mainLoop2 (X0, Y0, V0, W, pre Z, Mn)
              else mainLoop (pre X, pre Y, pre V, W, Mn);
  -- coda
  Z = coda (X, Y, V, W, S, T);
tel;
\end{lstlisting}
\end{minipage}
 \caption{Textual representation of the MAC node}
 \label{fig:MAC}
\end{figure*}
Keys could also be represented using the type Pair, but we prefer introducing the following dedicated structure \lstinline+Key+ to clearly distinguish between keys and other pairs (e.g., results of the multiplication of two blocks):
\begin{lstlisting}[language=Lustre]
type Key = struct {K: Block; J: Block};
\end{lstlisting}
% main functions
We define the ``multiplicative'' functions used for MAA computations, most of which are presented in~\cite{Davies-Clayden-88} or have been later introduced in~\cite{Menezes-vanOorschot-Vanstone-96}.
The three principal low-level operations are MUL1, MUL2, and MUL2A, with its auxiliary functions.
We also define the higher-level functions that implement the MAA algorithm on one segment (maximum 1024-byte), namely prelude, inner loop, coda, as well as the principal function MAC (that computes the 4-byte signature of a message).

%\subsection{Modelling choices}
It turns out that SCADE/Lustre enables an elegant modeling of the involved data structures, such as enumeration (e.g., the enumeration \lstinline[language=Lustre]+Bit+) and structures (e.g., the type Byte defined as \lstinline[language=Lustre]+Octet+).
We briefly discuss below some of our choices for modeling the MAA in SCADE/Lustre.
\begin{description}
\item[Usage of local variables.]
Local variables are essential to store computed results that need to be used several times, thus avoiding identical calculations to be repeated.
SCADE allows us to freely define and assign local variables; the compiler guarantees that each variable is duly assigned before used.
However, SCADE forbids successive assignments to the same variable. 
For instance, the \lstinline[language=Lustre]+MUL2+ function can be expressed in SCADE as follows:
\begin{lstlisting}[language=Lustre]
function MUL2 (w1, w2 : Block) returns (w: Block)
var w1w2, w3w4, w5w6: Pair; w3: Block;
let
  w1w2 = mulBlock (w1, w2);
  w3w4 = ADDC (w1w2.w1, w1w2.w1);
  w3 = addBlock (w3w4.w2, addBlock (w3w4.w1, w3w4.w1));
  w5w6 = ADDC (w3, w1w2.w2);
  w = addBlock (w5w6.w2, addBlock (w5w6.w1, w5w6.w1));
tel;
\end{lstlisting}
\item[Functions computing several results.]
There are several such functions in the MAA; let us consider the \lstinline[language=Lustre]+prelude+ function which takes two block parameters \lstinline[language=Lustre]+J+ and \lstinline[language=Lustre]+K+ and returns six block parameters \lstinline[language=Lustre]+X+, \lstinline[language=Lustre]+Y+, \lstinline[language=Lustre]+V+, \lstinline[language=Lustre]+W+, \lstinline[language=Lustre]+S+, and \lstinline[language=Lustre]+T+.
By exploiting the fact that SCADE/Lustre functions may return one or several parameters, we can elegantly define the \lstinline[language=Lustre]+prelude+ function in SCADE as follows:
\begin{lstlisting}[language=Lustre]
function prelude (J, K: Block) 
         returns (X, Y, V, W, S, T: Block)
var P: Octet; H4, H5, H6, H7, H8, H9: Block;
    J1, J14, J16, J18, J24, J26, J28: Block;
    K1, K15, K17, K25, K27, K19, K29: Block;
let
  J1, K1 = BYT (J, K);
  P = PAT (J, K);
  _,J14, J16, J18,_,J24, J26, J28 = preludeJ (J1);
  _,_,K15, K17, K19,_,_,K25, K27, K29 = preludeK (K1);
  H4, H6, H8 = preludeHJ (J14,J16,J18,J24,J26,J28);
  _,H5, H7, H9 = preludeHK (K15,K17,K19,K25,K27,K29,P);
  X, Y = BYT (H4, H5);
  V, W = BYT (H6, H7);
  S, T = BYT (H8, H9);
tel;
\end{lstlisting}
The definitions of the auxiliary functions \lstinline[language=Lustre]+preludeJ+, \lstinline[language=Lustre]+preludeK+, \lstinline[language=Lustre]+preludeHJ+, and \lstinline[language=Lustre]+preludeHK+ are in Appendix~\ref{sec:maa-mai-funct}.
This function is invoked as follows (as it is done in the MAC node):
\begin{lstlisting}[language=Lustre]
X, Y, V, W, S, T = prelude (J, K);
\end{lstlisting}
Note that in SCADE the underscore symbol (\lstinline[language=Lustre]+_+) is used in place of defining a local variable if the corresponding returned value is not used.
In our case, we needed some outputs from the auxiliary functions for the test vectors, but not for the \lstinline[language=Lustre]+prelude+ function, which is why we 'assigned them to \lstinline[language=Lustre]+_+' in \lstinline[language=Lustre]+prelude+.
\end{description}

\paragraph{Comparison between SCADE and Lustre implementations:}
In the semantic context of our MAA formal models the differences between the two languages are negligible.
First, SCADE uses comma instead of a semicolon, and defines the type before the equal in the constant/function definitions.
Second, using let/tel in a function composed of one statement is not permitted in SCADE, which generally reduces the number of code lines.
However, the main difference is on the compilers: the SCADE compiler enabled us to detect some potentially non-initialized variables in the \lstinline[language=Lustre]+MAC+ node and variables not used in the \lstinline[language=Lustre]+prelude+ function.
The Lustre and SCADE models contain 9 types (structures), 442 constants, 100 functions, and 1 nodes each.
Our complete Lustre model of the MAA (see the repository) consists of 1908 lines and our complete SCADE model of the MAA (see Appendix~\ref{ap:maa}) has 1762 lines and 3 additional graphical nodes.

\paragraph{Comparison between SCADE/Lustre and earlier implementations:}
% executable
Our two synchronous models are executable, in a sense that they can be compiled automatically into some executable program that will be run to generate the expected results.
Let us review these two synchronous models with the three earlier executable specifications; REC-17~\cite{Garavel-Marsso-17}, LOTOS-18~\cite{Garavel-Marsso-18}, and, LNT-18~\cite{Garavel-Marsso-18}.
\begin{itemize}
% performance + full contain + formal (cf mistakes found in external code LNT)
\item Rather than importing external fragments of code written in C, as in LOTOS-18 and LNT-17, to implement for instance blocks and arithmetic operations on blocks, REC-17 and Lustre/SCADE models implement the algorithm entirely, they are self-contained and fully-formal.
Even if using external fragments would allow a shorter formalization and generation of more efficient code.
We prefer to avoid the risks arising when formal and non-formal codes are mixed (e.g., an error was found in the external implementation in C~\cite{Garavel-Marsso-18}).
% original application, scade is a big public
\item Contrary to REC~17, LNT-18 and LOTOS-18 specifications, the Lustre/SCADE models contains duplicated computations.
Effectively, the MAC node is defined as a sequential function; at each instant, the MAC calculates prelude and coda, even if prelude is only needed at the beginning of each message sequence, and the latter is only needed at the end.
Such ``trick'' was necessary, since modelling cryptographical protocol is not the usual application domain of Lustre/SCADE.
\item SCADE and LNT-18 models have a more disciplined specification style than the three other executable specifications, because of the numerous static-analysis checks (e.g., unused variables, useless assignments, etc.) performed by their compilers.
\item Finally, the SCADE model contains graphical representations, which makes the model readable and easier to use for non-specialists.
To the best of our knowledge, this feature is absent in other specification languages.
\end{itemize}
% ----------------------------------------------------------------------
%  model validation

\section{Testing the MAA Model}
\label{sec:validation}
% unit tests on the functions
To validate our SCADE and Lustre models, we define four sets of test vectors derived from the specification in~\cite{Davies-Clayden-88}. The four sets are defined as follows:
\begin{enumerate}[(T1)]
  \item implements the 36 checks listed in Tables 1, 2, and 3 of~\cite{Davies-Clayden-88}.
  These test vectors specify, for a few given keys and messages, the expected values of intermediate calculations (e.g., MUL1, MUL2, MUL2A, etc.).
  \item is based upon Table 4 of~\cite{Davies-Clayden-88}, checking if the main loop of the MAA (as described on page 10 of~\cite{Davies-Clayden-88}) is correctly implemented on six groups of checks (three single-block messages and one three-block message); T2 corresponds to 56 tests.
  \item is based upon Table 5 of~\cite{Davies-Clayden-88}, checking if the MAA signature is correctly computed on  four groups of checks, with two different keys and two different messages; T3 corresponds to 64 tests. 
  \item checks all intermediary values of the algorithm with a message of 20 blocks containing only zeros directly taken from Table 6 of~\cite{Davies-Clayden-88}; T4 corresponds to 45 tests.
\end{enumerate}
We automate the test execution process, for the test vector (T3) presented in~\cite[Table 6]{Davies-Clayden-88}, namely, four groups of checks, with two different keys and two different messages.
The test automation process is different for Lustre and SCADE; we explain them in the following section.
% figure of Abstract test cases generated
%~ \begin{figure}[h]
  %~ \centering
%~ \begin{tabular}{|c|c|c|c|c|}
  %~ \hline
  %~ MAA Variables & Message 1 & Message 2 & Message 3 &  Message 4\\
  %~ \hline
  %~ J & 00FF 00FF & 00FF 00FF & 5555 5555 & 5555 5555\\ 
  %~ K & 0000 0000 & 0000 0000  & 5A35 D667 & 5A35 D667\\ 
  %~ P & FF & FF & 00 & 00\\ 
  %~ X0 & 4A64 5A01 & 4A64 5A01 & 34AC F886 & 34AC F886\\ 
  %~ Y0 & 50DE C930 & 50DE C930 & 7397 C9AE & 7397 C9AE\\ 
  %~ V0 & 5CCA 3239 & 5CCA 3239 & 7201 F4DC & 7201 F4DC\\ 
  %~ W & FECC AA6E & FECC AA6E & 2829 040B & 2829 040B\\ 
  %~ M1 & 5555 5555 & AAAA AAAA & 0000 0000 & FFFF FFFF\\ 
  %~ X & 48B2 04D6 & 6AEB ACF8 & 2FD7 6FFB & 8DC8 BBDE\\ 
  %~ Y & 5834 A585 & 9DB1 5CF6 & 550D 91CE & FE4E 5BDD\\ 
  %~ M2 & AAAA AAAA & 5555 5555 & FFFF FFFF & 0000 0000\\ 
  %~ X & 4F99 8E01 & 270E EDAF & A70F C148 & CBC8 65BA\\ 
  %~ Y & BE9F 0917 & B814 2629 & 1D10 D8D3 & 0297 AF6F\\ 
  %~ S & 51ED E9C7 & 51ED E9C7 & 9E2E 7B36 & 9E2E 7B36\\ 
  %~ X & 3449 25FC & 2990 7CD8 & B1CC 1CC5 & 3CF3 A7D2\\ 
  %~ Y & DB91 02B0 & BA92 DB12 & 29C1 485F & 160E E9B5\\ 
  %~ T & 24B6 6FB5 & 24B6 6FB5 & 1364 7149 & 1364 7149\\ 
  %~ X & 277B 4B25 & 28EA D8B3 & 288F C786 & D048 2465\\ 
  %~ Y & D636 250D & 81D1 0CA3 & 9115 A558 & 7050 EC5E\\ 
  %~ Z & F14D 6E28 & A93B D410 & B99A 62DE & A018 C83B\\
  %~ \hline
%~ \end{tabular}
%~ \caption{Tests (T3) inspired by Table 5 of~\cite{Davies-Clayden-88}.}
%~ \label{fig:t4vectors}
%~ \end{figure}
%TODO: Sentence telling we have two validations

\subsection{Validation of the Lustre model}
% init tests on the whole specification
When automating the test execution process, we use the testing tool Lurette~\cite{Jahier-Raymond-Baufreton-06} and take advantage of its connection with the Lustre language.
% FIGURE LURETTE WITH MAA
\begin{figure}[h]
  \centering
  %\Figure{luretteMaa}
  \includegraphics[scale=0.15]{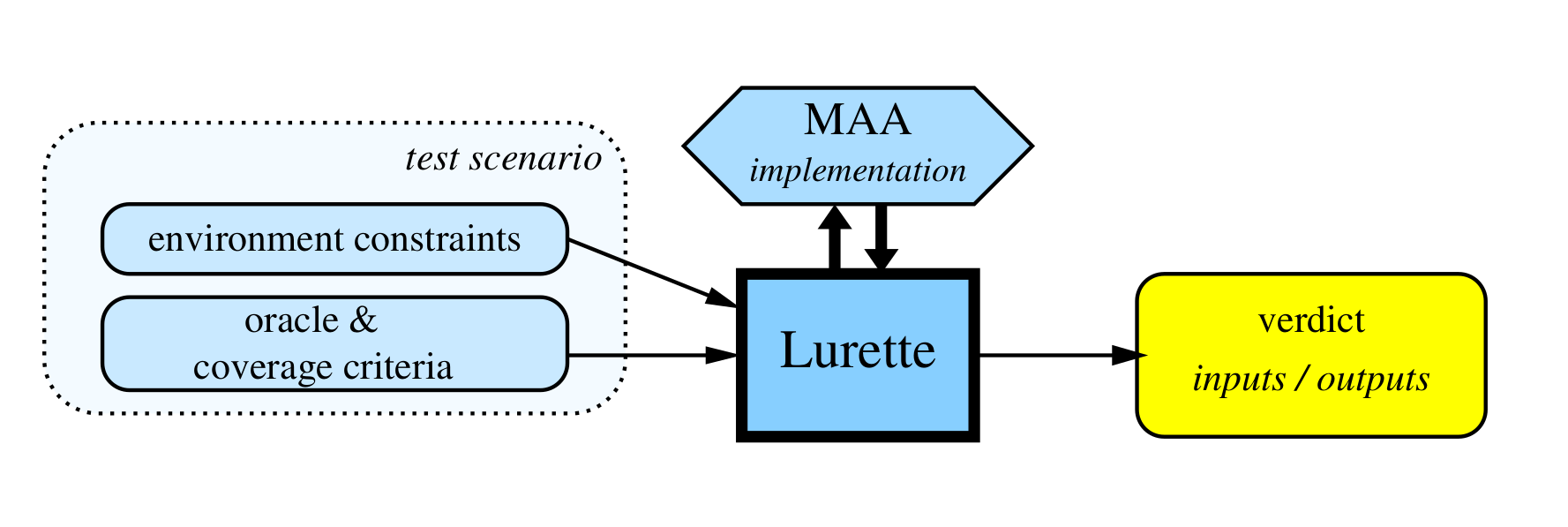}
  \caption{Overview of the Lurette testing tool}
  \label{fig:luretteMaa}
\end{figure}
Figure~\ref{fig:luretteMaa} gives an overview of Lurette in the context of testing our MAA model.
Lurette takes two inputs:
(i) a specification of the environment constraints in Lutin~\cite{Raymond-Roux-Jahier-08} to dynamically constrain the inputs,
(ii) an oracle implementing the test decision, in our case the 12 possible pairs (one key K J, and a message of two blocks), and the expected resulting MAC; 
and some parameters controlling the execution and the coverage of the generated input sequences, e.g., the number of steps (n) in the execution sequence.
Lurette interacts with the Lustre implementation of the MAA, logs the generated sequence of inputs and their corresponding outputs in a file, and displays the test decisions.

Following~\cite{Garavel-Marsso-17}, we add three supplementary test vectors that operate on messages of 16, 256, and 4100 blocks containing bit patterns not preserved by permutations to detect implementation mistakes arising from byte permutations (e.g., endianness issues) or incorrect segmentation of messages longer than 1024 bytes (i.e., 256 blocks).
Originally in~\cite{Garavel-Marsso-17}, it was done by introducing a function acting as a pseudo-random message generator, which is only possible to implement in Lustre with recursive nodes.
Thus, we rewrite the node \lstinline+MAC+ as the recursive Lustre \lstinline+MAC_long_messages+ node as below:
%~ \begin{lstlisting}[language=Lustre]
%~ node MAC (id: int; init: bool) 
         %~ returns (X, Y, V, W, S, T, Z: Block; n: int);
%~ var X0, Y0, V0, KJ: Key; Mn: Block;
%~ let
  %~ Mn, KJ = get_mess_key (id); ...
%~ \end{lstlisting}
\begin{lstlisting}[language=Lustre]
node MAC_long_messages <<const it : int >> (INIT, INCR: Block;KJ: Key) 
                         returns (mac: Block);
var X0, Y0, V0: Block; 
    X, Y, V, W, S, T, Z: Block;
    Mn: Block; n: int; newSegment: bool;
let
  -- pseudo-random message generation
  Mn = INIT -> addBlock (pre Mn, INCR); 
  -- ... identical to the MAC node
  -- recursive call on the next block
  mac = with (it = 1) then Z
        else MAC_long_messages << it-1 >> (INIT, INCR, KJ);
tel;
\end{lstlisting}
When defining the recursive node in the code above, we specify the integer depth of recursion in triangular brackets, the depth of the recursion.
The \lstinline+MAC_long_messages+ has three inputs: the blocks \lstinline+INIT+ and \lstinline+INCR+ to generates a message of the given length (\lstinline+it+), the blocks of which follow an arithmetic progression, and the key \lstinline+KJ+.

The following Lustre node \lstinline+check_long_message+ implements the test vectors that operate on messages of 16, 256, and 4100 blocks containing bit patterns not preserved by permutations:
%~ \begin{lstlisting}[language=Lustre]
%~ node environment () returns (id: int; init: bool) =
%~ loop {
  %~ |id = 1 and init = true fby id = 12 and init = false
  %~ |id = 2 and init = true fby id = 22 and init = false
  %~ |id = 3 and init = true fby id = 32 and init = false
  %~ |id = 4 and init = true fby id = 42 and init = false } 
%~ \end{lstlisting}
\begin{lstlisting}[language=Lustre]
node check_long_message () returns (v: bool);
var m20, m16, m256, m4100: Block; KJ: Key;
let
  KJ = Key {K = x80018000; J = x80018001};
  
  -- messages of 16, 256, and 4100 blocks
  m16 = MAC_long_messages<<16>>(x00000000, x07050301, KJ);
  m20 = MAC_long_messages<<20>>(x00000000, x00000000, KJ);
  m256 = MAC_long_messages<<256>>(x00000000, x07050301, KJ);
  m4100 = MAC_long_messages<<4100>>(x00000000, x07050301, KJ);
  
  v =  (m20 = xDB79FBDC) and (m16 = x8CE37709) and
       (m256 = x717153D5) and (m4100 = x7783C51D);
tel;
\end{lstlisting}
Note that these recursive nodes are not available in SCADE.
Thus, we did not implement the three supplementary test vectors~\cite{Garavel-Marsso-17} for the SCADE model. 

The Lutin environment constrains the input of the MAA to one out of four messages, each one containing two words.
Note that if the input values are not explicitly constrained in the Lutin environment, random numbers will be generated.
The environment contains the 12 possible pairs of keys and messages.
Although larger and more complex environments could be written, this task is tedious and error-prone, in particular due to the representation of messages and keys by natural numbers.
In our example, correct tests consist of a correct MAC value, given one key and a message.
A small excerpt of oracle for our Lustre model is given by the following Lustre code, containing the test decisions and the expected output: the auxiliary block values of X, Y and MAC value for each message block given in the first columnn of~\cite[Table 6]{Davies-Clayden-88}.
\begin{lstlisting}[language=Lustre]
node oracle (id: int; init: bool; X, Y, Z: Block; n: int) 
             returns (res: bool);
let
  res = true -> 
   ((id = 1 and init and X = x48B204D6 and Y = x5834A585) or
    (id = 2 and init and X = x6AEBACF8 and Y = x9DB15CF6) or ...
    and CHECK () = true and check_long_message () = true);
tel;
\end{lstlisting}
The whole Lutin scenario and Lustre oracle for testing the Lustre model with the test vector (T3) will appear in the MARS Workshop repository.

% Statistiques
The Lustre and the test scenarios for the sets of test vectors (T1), (T2), (T3), (T4) has 1908 lines and contains 9 types (structures), 442 constants, 100 functions, and 4 nodes.
Our Lustre model of the MAA was successfully validated by the 203 tests.

\subsection{Validation of the SCADE model}
We automate the test execution process, by using the SCADE test environment toolset.
To achieve this goal, one has to write a test scenario in an SSS file, with the description of the test sequences for testing the main MAC node.
The scenario sets the input values and defines the expected output values to be checked.
For instance, for our SCADE MAA model, expected output values consist of the correct MAC value (Z) or the three intermediate values (X, Y, and V), given one key and a message as inputs.
A small excerpt of the test scenario is given by the following SCADE code, containing the input sequences: a key, a message block and a boolean indicating if it is the first block of the message. The expected outputs are the auxiliary block values of X, Y and MAC value for each block of message given in~\cite[Table 6]{Davies-Clayden-88}:
\begin{lstlisting}[language=Lustre]
## -- inputs --------------------------------------
# -- Key
SSM::set MAC/KJ 
     (((X0,X0,X0,X0,X0,X0,X0,X0),(X0,X0,X0,X0,X0,X0,X0,X0),
       (X0,X0,X0,X0,X0,X0,X0,X0),(X0,X0,X0,X0,X0,X0,X0,X0)),
      ((X0,X0,X0,X0,X0,X0,X0,X0),(X1,X1,X1,X1,X1,X1,X1,X1),
       (X0,X0,X0,X0,X0,X0,X0,X0),(X1,X1,X1,X1,X1,X1,X1,X1)))
# -- Block
SSM::set MAC/Mn 
     ((X0,X1,X0,X1,X0,X1,X0,X1),(X0,X1,X0,X1,X0,X1,X0,X1),
      (X0,X1,X0,X1,X0,X1,X0,X1),(X0,X1,X0,X1,X0,X1,X0,X1))
SSM::set MAC/init true
## -- outputs --------------------------------------
# -- Block 
SSM::check MAC/X 
    ((X0,X1,X0,X0,X1,X0,X0,X0),(X1,X0,X1,X1,X0,X0,X1,X0),
     (X0,X0,X0,X0,X0,X1,X0,X0),(X1,X1,X0,X1,X0,X1,X1,X0))
SSM::check MAC/Y 
    ((X0,X1,X0,X1,X1,X0,X0,X0),(X0,X0,X1,X1,X0,X1,X0,X0),
     (X1,X0,X1,X0,X0,X1,X0,X1),(X1,X0,X0,X0,X0,X1,X0,X1))
# -- intermediate value
SSM::uncheck MAC/Z
## ----------------------------------------------
SSM::cycle ...
\end{lstlisting}
The SCADE scenario files (.sss) contain the following instructions:
\begin{itemize}
\item \lstinline[language=Lustre]+SSM::set+: assign a specific value to an input for the next execution cycle;
\item \lstinline[language=Lustre]+SSM::check+: register a test of an output against a specified value;
\item \lstinline[language=Lustre]+SSM::uncheck+: unregister the active test for an observable output;
\item \lstinline[language=Lustre]+SSM::cycle+: set the number of execution cycles in a test sequence (the default number of cycles is set to 1 when no argument is specified);
\item \lstinline[language=Lustre]+#+: comment the line.
\end{itemize}
Note that tests are registered by the next calls to \lstinline[language=Lustre]+SSM::cycle+ instruction.
The whole scenario for testing our SCADE model with the test vector (T3) will appear in the MARS Workshop repository.
The test vector with long messages (i.e., containing more than two blocks) and the test vectors for the auxiliary functions are hard coded in functions, as for instance, the twenty-words message in the Appendix~\ref{ap:20words}.

% Statistiques
The SCADE model and the test scenarios for the sets of test vectors (T1), (T2), (T3), (T4), which is 201 tests in total, have 1894 lines and contain 9 types (structures), 442 constants, 99 functions, and 2 nodes.
Our SCADE model of the MAA was successfully validated by these 201 tests.

\section{Conclusion and Future Work}
\label{sec:conc}

% summary
In this paper we have presented a formal SCADE model of the Message Authenticator Algorithm (MAA).
First, we gave an overview of this cryptographic functions for computing a Message Authentication Code, then we discussed the choices and the challenges of this implementation in SCADE and compared it to the Lustre implementation.
Finally we validated our model using the test test vectors derived from the specification in~\cite{Davies-Clayden-88} 

% mistakes
The modelling of the MAA and its validation enabled us to find three errors in the Lustre v6 toolbox: (i) lv6 compiler generates invalid C code if a variable identifier uses quotes ('); (ii) another issue with constant structures and the compiler lv6; and (iii) Lurette tool ignoring -2c-exec option; all these errors have been communicated to the Lustre v6 developers and have been quickly fixed by the team.
The modelling of the SCADE model enabled us to improve the Lustre model, more precisely by detecting potentially non-initialized variables and unused variables. 

% manual testing is tedious
Since neither SCADE, nor Lustre implements the list data type, we had to consider only messages with fixed size in our SCADE model. However, in our Lustre model, we could generate pseudo-random messages by adding a recursive node.
It is however fair to warn the reader that testing our Lustre/SCADE MAA models was a tedious task, i.e., we had to manually define more than 400 constants.

% future work?
In this paper we only have written the main node computing the MAC with the graphical language SCADE.
As future work, the full MAA algorithm or at least its main functions could be designed with the graphical language SCADE and some test scenarios could also be simulated on the graphical nodes.
To improve the code efficiency, one could as well implement blocks and arithmetic operations on blocks in external fragments of code written in C, and import them in the Lustre/SCADE models.

%\Sectionh{Acknowledgment}
%We are grateful to Hubert Garavel and Wendelin Serwe for helpful remarks about the paper and the LNT model.
%\nocite{*}
\bibliographystyle{eptcs}
\bibliography{biblio}

% ------------------------------------------------------------------------------
% Appendices
\clearpage
\appendix
\section{Formal Model of the MAA in SCADE}
\label{ap:maa}
\input{maascade}
\end{document}

%% file: convecs-3.tex
\usepackage[utf8]{inputenc} % Linux

\usepackage[OT1]{fontenc}

\usepackage{fancyhdr} % nouveau nom de fancyheadings
\usepackage{makeidx}
\usepackage{graphicx}
\usepackage{color}
\usepackage{array}

\ifx\hypersetup\undefined
\usepackage[hyperfootnotes=false]{hyperref}
\fi

\usepackage{breakurl} % indispensable pour couper proprement les URLs longues
\usepackage{amssymb} % for \varnothing
\usepackage{amsfonts}
\ifx\keywords\undefined
\newcommand{\keywords}[1]{\begin{quote}{\bf Keywords:} #1\end{quote}}
\fi

\definecolor{grey}{cmyk}{0,0,0,0.7}

\definecolor{blue}{cmyk}{.6,0.1,0,0}
\definecolor{tblue}{rgb}{0,0,.7}
\definecolor{red}{cmyk}{0,.6,0.6,0}
\definecolor{green}{cmyk}{0.5,0,0.7,0}
\definecolor{magenta}{cmyk}{0,.8,0,0}
\definecolor{purple}{rgb}{.8,0,1}
\definecolor{lpurple}{cmyk}{.2,.6,0,0}
\definecolor{pink}{rgb}{1,.6,.6}
\definecolor{gold}{cmyk}{0,.1,.7,0}
\definecolor{orange}{cmyk}{0,.3,.6,0}
\definecolor{goldorange}{cmyk}{0,.2,.7,0}

\hypersetup{
pdfborder= 0 0 0,
colorlinks=true,
anchorcolor=grey,
citecolor=grey,
filecolor=grey,
linkcolor=grey,
menucolor=grey,
runcolor=grey,
urlcolor=grey
}

\renewcommand{\epsilon}{\varepsilon}
\renewcommand{\phi}{\varphi}

\ifx\And\undefined
\newcommand{\And}{\wedge}
\fi

\ifx\AND\undefined
\newcommand{\AND}{\bigwedge}
\fi

\newcommand{\onlymath}{\null\ifmmode\else\errmessage{Your macro should only be used in math-mode}\fi}
\newcommand{\IGNORE}[1]{}

%% file: macr_pic.tex
\input{psfig}

\newlength{\magic}

\makeatletter
\def\Picture[#1]{\@ifnextchar[{\@lPicture[#1]}{\@Picture[#1]}}
\def\@Picture[#1]#2{\par\begin{center}\@lPicture[#1][14]{#2}\end{center}\par}
\def\@lPicture[#1][#2]#3{\mbox{\vspace*{1ex}\hspace*{0.8cm}\psfig{rheight=#1cm,rwidth=#2cm,bbllx=3.1714cm,bblly=0cm,bburx=0cm,bbury=\magic,figure=#3.ps}\hspace*{-0.8cm}\vspace*{1ex}}}
\makeatother

\makeatletter
\def\Figure{\@ifnextchar[{\@lFigure}{\@Figure}}
\def\@Figure#1{\begin{center} \input{#1} \end{center}}
\def\@lFigure[#1]#2{\hspace*{#1}{\parbox{\textwidth}{\@Figure{#2}}}}
\makeatother

%% file: maascade.tex
This annex presents the specification of the MAA in the SCADE language.
This specification is fully self-contained, meaning that it does not depend on any externally-defined library -- with the minor disadvantage of somewhat lengthy definitions for byte and blocks constants.
For readability, the specification has been split into 16 parts, each part being devoted to a particular type, a group of functions sharing a common purpose, or a collection of test vectors. 
The first parts contain general definitions that are largely independent from the MAA; starting from Section~\ref{section:maa-related}, the definitions become increasingly more MAA-specific.
All machine words (bytes, blocks, etc.) are represented according to the ``big endian'' convention, i.e., the first argument of each corresponding constructor denote the most significant bit.

\section{Definitions for type Bit}
We define bits using an enumeration \lstinline+Bit+ (\lstinline+X0+ and \lstinline+X1+), together with functions implementing logical operations on bits.
\begin{lstlisting}[language=Lustre]
------------------------------------------
type Bit = enum {X0, X1};
------------------------------------------
function notBit (x1: Bit) returns (x: Bit)
  x = if x1 = X0 then X1 else X0;
------------------------------------------
function andBit (x1, x2: Bit) returns (x: Bit)
  x = if x2 = X0 then X0 else x1;
------------------------------------------
function orBit (x1, x2: Bit) returns (x: Bit)
  x = if x2 = X0 then x1 else X1;
------------------------------------------
function xorBit (x1, x2: Bit) returns (x: Bit)
  x = if x2 = X0 then x1 else notBit (x1);
------------------------------------------
\end{lstlisting}
\section{Definitions for Type Byte}
We define bytes (octets) using a structure \lstinline+Octet+ that contains the eight bits of a byte, together with functions implementing bitwise logical operations, left-shift and right-shift operations on bytes, as well as all byte constants needed to formally describe the MAA and its test vectors.
\begin{lstlisting}[language=Lustre]
------------------------------------------
type Octet = {x1: Bit, x2: Bit, x3: Bit, x4: Bit,
              x5: Bit, x6: Bit, x7: Bit, x8: Bit};
------------------------------------------
const 
  x00:Octet = {x1: X0, x2: X0, x3: X0, x4: X0,
               x5: X0, x6: X0, x7: X0, x8: X0};
  x01:Octet = {x1: X0, x2: X0, x3: X0, x4: X0,
               x5: X0, x6: X0, x7: X0, x8: X1};
  x02:Octet = {x1: X0, x2: X0, x3: X0, x4: X0,
               x5: X0, x6: X0, x7: X1, x8: X0};
  x03:Octet = {x1: X0, x2: X0, x3: X0, x4: X0,
               x5: X0, x6: X0, x7: X1, x8: X1};
  x04:Octet = {x1: X0, x2: X0, x3: X0, x4: X0,
               x5: X0, x6: X1, x7: X0, x8: X0};
  x05:Octet = {x1: X0, x2: X0, x3: X0, x4: X0,
               x5: X0, x6: X1, x7: X0, x8: X1};
  x06:Octet = {x1: X0, x2: X0, x3: X0, x4: X0,
               x5: X0, x6: X1, x7: X1, x8: X0};
  x07:Octet = {x1: X0, x2: X0, x3: X0, x4: X0,
               x5: X0, x6: X1, x7: X1, x8: X1};
  x08:Octet = {x1: X0, x2: X0, x3: X0, x4: X0,
               x5: X1, x6: X0, x7: X0, x8: X0};
  x09:Octet = {x1: X0, x2: X0, x3: X0, x4: X0,
               x5: X1, x6: X0, x7: X0, x8: X1};
  x0A:Octet = {x1: X0, x2: X0, x3: X0, x4: X0,
               x5: X1, x6: X0, x7: X1, x8: X0};
  x0B:Octet = {x1: X0, x2: X0, x3: X0, x4: X0,
               x5: X1, x6: X0, x7: X1, x8: X1};
  x0C:Octet = {x1: X0, x2: X0, x3: X0, x4: X0,
               x5: X1, x6: X1, x7: X0, x8: X0};
  x0D:Octet = {x1: X0, x2: X0, x3: X0, x4: X0,
               x5: X1, x6: X1, x7: X0, x8: X1};
  x0E:Octet = {x1: X0, x2: X0, x3: X0, x4: X0,
               x5: X1, x6: X1, x7: X1, x8: X0};
  x0F:Octet = {x1: X0, x2: X0, x3: X0, x4: X0,
               x5: X1, x6: X1, x7: X1, x8: X1};
  x10:Octet = {x1: X0, x2: X0, x3: X0, x4: X1,
               x5: X0, x6: X0, x7: X0, x8: X0};
  x11:Octet = {x1: X0, x2: X0, x3: X0, x4: X1,
               x5: X0, x6: X0, x7: X0, x8: X1};
  x12:Octet = {x1: X0, x2: X0, x3: X0, x4: X1,
               x5: X0, x6: X0, x7: X1, x8: X0};
  x13:Octet = {x1: X0, x2: X0, x3: X0, x4: X1,
               x5: X0, x6: X0, x7: X1, x8: X1};
  x14:Octet = {x1: X0, x2: X0, x3: X0, x4: X1,
               x5: X0, x6: X1, x7: X0, x8: X0};
  x15:Octet = {x1: X0, x2: X0, x3: X0, x4: X1,
               x5: X0, x6: X1, x7: X0, x8: X1};
  x16:Octet = {x1: X0, x2: X0, x3: X0, x4: X1,
               x5: X0, x6: X1, x7: X1, x8: X0};
  x17:Octet = {x1: X0, x2: X0, x3: X0, x4: X1,
               x5: X0, x6: X1, x7: X1, x8: X1};
  x18:Octet = {x1: X0, x2: X0, x3: X0, x4: X1,
               x5: X1, x6: X0, x7: X0, x8: X0};
  x1A:Octet = {x1: X0, x2: X0, x3: X0, x4: X1,
               x5: X1, x6: X0, x7: X1, x8: X0};
  x1B:Octet = {x1: X0, x2: X0, x3: X0, x4: X1,
               x5: X1, x6: X0, x7: X1, x8: X1};
  x1C:Octet = {x1: X0, x2: X0, x3: X0, x4: X1,
               x5: X1, x6: X1, x7: X0, x8: X0};
  x1D:Octet = {x1: X0, x2: X0, x3: X0, x4: X1,
               x5: X1, x6: X1, x7: X0, x8: X1};
  x1E:Octet = {x1: X0, x2: X0, x3: X0, x4: X1,
               x5: X1, x6: X1, x7: X1, x8: X0};
  x1F:Octet = {x1: X0, x2: X0, x3: X0, x4: X1,
               x5: X1, x6: X1, x7: X1, x8: X1};
  x20:Octet = {x1: X0, x2: X0, x3: X1, x4: X0,
               x5: X0, x6: X0, x7: X0, x8: X0};
  x21:Octet = {x1: X0, x2: X0, x3: X1, x4: X0,
               x5: X0, x6: X0, x7: X0, x8: X1};
  x22:Octet = {x1: X0, x2: X0, x3: X1, x4: X0,
               x5: X0, x6: X0, x7: X1, x8: X0};
  x23:Octet = {x1: X0, x2: X0, x3: X1, x4: X0,
               x5: X0, x6: X0, x7: X1, x8: X1};
  x24:Octet = {x1: X0, x2: X0, x3: X1, x4: X0,
               x5: X0, x6: X1, x7: X0, x8: X0};
  x25:Octet = {x1: X0, x2: X0, x3: X1, x4: X0,
               x5: X0, x6: X1, x7: X0, x8: X1};
  x26:Octet = {x1: X0, x2: X0, x3: X1, x4: X0,
               x5: X0, x6: X1, x7: X1, x8: X0};
  x27:Octet = {x1: X0, x2: X0, x3: X1, x4: X0,
               x5: X0, x6: X1, x7: X1, x8: X1};
  x28:Octet = {x1: X0, x2: X0, x3: X1, x4: X0,
               x5: X1, x6: X0, x7: X0, x8: X0};
  x29:Octet = {x1: X0, x2: X0, x3: X1, x4: X0,
               x5: X1, x6: X0, x7: X0, x8: X1};
  x2A:Octet = {x1: X0, x2: X0, x3: X1, x4: X0,
               x5: X1, x6: X0, x7: X1, x8: X0};
  x2B:Octet = {x1: X0, x2: X0, x3: X1, x4: X0,
               x5: X1, x6: X0, x7: X1, x8: X1};
  x2D:Octet = {x1: X0, x2: X0, x3: X1, x4: X0,
               x5: X1, x6: X1, x7: X0, x8: X1};
  x2E:Octet = {x1: X0, x2: X0, x3: X1, x4: X0,
               x5: X1, x6: X1, x7: X1, x8: X0};
  x2F:Octet = {x1: X0, x2: X0, x3: X1, x4: X0,
               x5: X1, x6: X1, x7: X1, x8: X1};
  x30:Octet = {x1: X0, x2: X0, x3: X1, x4: X1,
               x5: X0, x6: X0, x7: X0, x8: X0};
  x31:Octet = {x1: X0, x2: X0, x3: X1, x4: X1,
               x5: X0, x6: X0, x7: X0, x8: X1};
  x32:Octet = {x1: X0, x2: X0, x3: X1, x4: X1,
               x5: X0, x6: X0, x7: X1, x8: X0};
  x33:Octet = {x1: X0, x2: X0, x3: X1, x4: X1,
               x5: X0, x6: X0, x7: X1, x8: X1};
  x34:Octet = {x1: X0, x2: X0, x3: X1, x4: X1,
               x5: X0, x6: X1, x7: X0, x8: X0};
  x35:Octet = {x1: X0, x2: X0, x3: X1, x4: X1,
               x5: X0, x6: X1, x7: X0, x8: X1};
  x36:Octet = {x1: X0, x2: X0, x3: X1, x4: X1,
               x5: X0, x6: X1, x7: X1, x8: X0};
  x37:Octet = {x1: X0, x2: X0, x3: X1, x4: X1,
               x5: X0, x6: X1, x7: X1, x8: X1};
  x38:Octet = {x1: X0, x2: X0, x3: X1, x4: X1,
               x5: X1, x6: X0, x7: X0, x8: X0};
  x39:Octet = {x1: X0, x2: X0, x3: X1, x4: X1,
               x5: X1, x6: X0, x7: X0, x8: X1};
  x3A:Octet = {x1: X0, x2: X0, x3: X1, x4: X1,
               x5: X1, x6: X0, x7: X1, x8: X0};
  x3B:Octet = {x1: X0, x2: X0, x3: X1, x4: X1,
               x5: X1, x6: X0, x7: X1, x8: X1};
  x3C:Octet = {x1: X0, x2: X0, x3: X1, x4: X1,
               x5: X1, x6: X1, x7: X0, x8: X0};
  x3D:Octet = {x1: X0, x2: X0, x3: X1, x4: X1,
               x5: X1, x6: X1, x7: X0, x8: X1};
  x3E:Octet = {x1: X0, x2: X0, x3: X1, x4: X1,
               x5: X1, x6: X1, x7: X1, x8: X0};
  x3F:Octet = {x1: X0, x2: X0, x3: X1, x4: X1,
               x5: X1, x6: X1, x7: X1, x8: X1};
  x40:Octet = {x1: X0, x2: X1, x3: X0, x4: X0,
               x5: X0, x6: X0, x7: X0, x8: X0};
  x41:Octet = {x1: X0, x2: X1, x3: X0, x4: X0,
               x5: X0, x6: X0, x7: X0, x8: X1};
  x42:Octet = {x1: X0, x2: X1, x3: X0, x4: X0,
               x5: X0, x6: X0, x7: X1, x8: X0};
  x43:Octet = {x1: X0, x2: X1, x3: X0, x4: X0,
               x5: X0, x6: X0, x7: X1, x8: X1};
  x44:Octet = {x1: X0, x2: X1, x3: X0, x4: X0,
               x5: X0, x6: X1, x7: X0, x8: X0};
  x45:Octet = {x1: X0, x2: X1, x3: X0, x4: X0,
               x5: X0, x6: X1, x7: X0, x8: X1};
  x46:Octet = {x1: X0, x2: X1, x3: X0, x4: X0,
               x5: X0, x6: X1, x7: X1, x8: X0};
  x47:Octet = {x1: X0, x2: X1, x3: X0, x4: X0,
               x5: X0, x6: X1, x7: X1, x8: X1};
  x48:Octet = {x1: X0, x2: X1, x3: X0, x4: X0,
               x5: X1, x6: X0, x7: X0, x8: X0};
  x49:Octet = {x1: X0, x2: X1, x3: X0, x4: X0,
               x5: X1, x6: X0, x7: X0, x8: X1};
  x4A:Octet = {x1: X0, x2: X1, x3: X0, x4: X0,
               x5: X1, x6: X0, x7: X1, x8: X0};
  x4B:Octet = {x1: X0, x2: X1, x3: X0, x4: X0,
               x5: X1, x6: X0, x7: X1, x8: X1};
  x4C:Octet = {x1: X0, x2: X1, x3: X0, x4: X0,
               x5: X1, x6: X1, x7: X0, x8: X0};
  x4D:Octet = {x1: X0, x2: X1, x3: X0, x4: X0,
               x5: X1, x6: X1, x7: X0, x8: X1};
  x4E:Octet = {x1: X0, x2: X1, x3: X0, x4: X0,
               x5: X1, x6: X1, x7: X1, x8: X0};
  x4F:Octet = {x1: X0, x2: X1, x3: X0, x4: X0,
               x5: X1, x6: X1, x7: X1, x8: X1};
  x50:Octet = {x1: X0, x2: X1, x3: X0, x4: X1,
               x5: X0, x6: X0, x7: X0, x8: X0};
  x51:Octet = {x1: X0, x2: X1, x3: X0, x4: X1,
               x5: X0, x6: X0, x7: X0, x8: X1};
  x53:Octet = {x1: X0, x2: X1, x3: X0, x4: X1,
               x5: X0, x6: X0, x7: X1, x8: X1};
  x54:Octet = {x1: X0, x2: X1, x3: X0, x4: X1,
               x5: X0, x6: X1, x7: X0, x8: X0};
  x55:Octet = {x1: X0, x2: X1, x3: X0, x4: X1,
               x5: X0, x6: X1, x7: X0, x8: X1};
  x58:Octet = {x1: X0, x2: X1, x3: X0, x4: X1,
               x5: X1, x6: X0, x7: X0, x8: X0};
  x5A:Octet = {x1: X0, x2: X1, x3: X0, x4: X1,
               x5: X1, x6: X0, x7: X1, x8: X0};
  x5B:Octet = {x1: X0, x2: X1, x3: X0, x4: X1,
               x5: X1, x6: X0, x7: X1, x8: X1};
  x5C:Octet = {x1: X0, x2: X1, x3: X0, x4: X1,
               x5: X1, x6: X1, x7: X0, x8: X0};
  x5D:Octet = {x1: X0, x2: X1, x3: X0, x4: X1,
               x5: X1, x6: X1, x7: X0, x8: X1};
  x5E:Octet = {x1: X0, x2: X1, x3: X0, x4: X1,
               x5: X1, x6: X1, x7: X1, x8: X0};
  x5F:Octet = {x1: X0, x2: X1, x3: X0, x4: X1,
               x5: X1, x6: X1, x7: X1, x8: X1};
  x60:Octet = {x1: X0, x2: X1, x3: X1, x4: X0,
               x5: X0, x6: X0, x7: X0, x8: X0};
  x61:Octet = {x1: X0, x2: X1, x3: X1, x4: X0,
               x5: X0, x6: X0, x7: X0, x8: X1};
  x62:Octet = {x1: X0, x2: X1, x3: X1, x4: X0,
               x5: X0, x6: X0, x7: X1, x8: X0};
  x63:Octet = {x1: X0, x2: X1, x3: X1, x4: X0,
               x5: X0, x6: X0, x7: X1, x8: X1};
  x64:Octet = {x1: X0, x2: X1, x3: X1, x4: X0,
               x5: X0, x6: X1, x7: X0, x8: X0};
  x65:Octet = {x1: X0, x2: X1, x3: X1, x4: X0,
               x5: X0, x6: X1, x7: X0, x8: X1};
  x66:Octet = {x1: X0, x2: X1, x3: X1, x4: X0,
               x5: X0, x6: X1, x7: X1, x8: X0};
  x67:Octet = {x1: X0, x2: X1, x3: X1, x4: X0,
               x5: X0, x6: X1, x7: X1, x8: X1};
  x69:Octet = {x1: X0, x2: X1, x3: X1, x4: X0,
               x5: X1, x6: X0, x7: X0, x8: X1};
  x6A:Octet = {x1: X0, x2: X1, x3: X1, x4: X0,
               x5: X1, x6: X0, x7: X1, x8: X0};
  x6B:Octet = {x1: X0, x2: X1, x3: X1, x4: X0,
               x5: X1, x6: X0, x7: X1, x8: X1};
  x6C:Octet = {x1: X0, x2: X1, x3: X1, x4: X0,
               x5: X1, x6: X1, x7: X0, x8: X0};
  x6D:Octet = {x1: X0, x2: X1, x3: X1, x4: X0,
               x5: X1, x6: X1, x7: X0, x8: X1};
  x6E:Octet = {x1: X0, x2: X1, x3: X1, x4: X0,
               x5: X1, x6: X1, x7: X1, x8: X0};
  x6F:Octet = {x1: X0, x2: X1, x3: X1, x4: X0,
               x5: X1, x6: X1, x7: X1, x8: X1};
  x70:Octet = {x1: X0, x2: X1, x3: X1, x4: X1,
               x5: X0, x6: X0, x7: X0, x8: X0};
  x71:Octet = {x1: X0, x2: X1, x3: X1, x4: X1,
               x5: X0, x6: X0, x7: X0, x8: X1};
  x72:Octet = {x1: X0, x2: X1, x3: X1, x4: X1,
               x5: X0, x6: X0, x7: X1, x8: X0};
  x73:Octet = {x1: X0, x2: X1, x3: X1, x4: X1,
               x5: X0, x6: X0, x7: X1, x8: X1};
  x74:Octet = {x1: X0, x2: X1, x3: X1, x4: X1,
               x5: X0, x6: X1, x7: X0, x8: X0};
  x75:Octet = {x1: X0, x2: X1, x3: X1, x4: X1,
               x5: X0, x6: X1, x7: X0, x8: X1};
  x76:Octet = {x1: X0, x2: X1, x3: X1, x4: X1,
               x5: X0, x6: X1, x7: X1, x8: X0};
  x77:Octet = {x1: X0, x2: X1, x3: X1, x4: X1,
               x5: X0, x6: X1, x7: X1, x8: X1};
  x78:Octet = {x1: X0, x2: X1, x3: X1, x4: X1,
               x5: X1, x6: X0, x7: X0, x8: X0};
  x79:Octet = {x1: X0, x2: X1, x3: X1, x4: X1,
               x5: X1, x6: X0, x7: X0, x8: X1};
  x7A:Octet = {x1: X0, x2: X1, x3: X1, x4: X1,
               x5: X1, x6: X0, x7: X1, x8: X0};
  x7B:Octet = {x1: X0, x2: X1, x3: X1, x4: X1,
               x5: X1, x6: X0, x7: X1, x8: X1};
  x7C:Octet = {x1: X0, x2: X1, x3: X1, x4: X1,
               x5: X1, x6: X1, x7: X0, x8: X0};
  x7D:Octet = {x1: X0, x2: X1, x3: X1, x4: X1,
               x5: X1, x6: X1, x7: X0, x8: X1};
  x7E:Octet = {x1: X0, x2: X1, x3: X1, x4: X1,
               x5: X1, x6: X1, x7: X1, x8: X0};
  x7F:Octet = {x1: X0, x2: X1, x3: X1, x4: X1,
               x5: X1, x6: X1, x7: X1, x8: X1};
  x80:Octet = {x1: X1, x2: X0, x3: X0, x4: X0,
               x5: X0, x6: X0, x7: X0, x8: X0};
  x81:Octet = {x1: X1, x2: X0, x3: X0, x4: X0,
               x5: X0, x6: X0, x7: X0, x8: X1};
  x83:Octet = {x1: X1, x2: X0, x3: X0, x4: X0,
               x5: X0, x6: X0, x7: X1, x8: X1};
  x84:Octet = {x1: X1, x2: X0, x3: X0, x4: X0,
               x5: X0, x6: X1, x7: X0, x8: X0};
  x85:Octet = {x1: X1, x2: X0, x3: X0, x4: X0,
               x5: X0, x6: X1, x7: X0, x8: X1};
  x86:Octet = {x1: X1, x2: X0, x3: X0, x4: X0,
               x5: X0, x6: X1, x7: X1, x8: X0};
  x89:Octet = {x1: X1, x2: X0, x3: X0, x4: X0,
               x5: X1, x6: X0, x7: X0, x8: X1};
  x8C:Octet = {x1: X1, x2: X0, x3: X0, x4: X0,
               x5: X1, x6: X1, x7: X0, x8: X0};
  x8D:Octet = {x1: X1, x2: X0, x3: X0, x4: X0,
               x5: X1, x6: X1, x7: X0, x8: X1};
  x8E:Octet = {x1: X1, x2: X0, x3: X0, x4: X0,
               x5: X1, x6: X1, x7: X1, x8: X0};
  x8F:Octet = {x1: X1, x2: X0, x3: X0, x4: X0,
               x5: X1, x6: X1, x7: X1, x8: X1};
  x90:Octet = {x1: X1, x2: X0, x3: X0, x4: X1,
               x5: X0, x6: X0, x7: X0, x8: X0};
  x91:Octet = {x1: X1, x2: X0, x3: X0, x4: X1,
               x5: X0, x6: X0, x7: X0, x8: X1};
  x92:Octet = {x1: X1, x2: X0, x3: X0, x4: X1,
               x5: X0, x6: X0, x7: X1, x8: X0};
  x93:Octet = {x1: X1, x2: X0, x3: X0, x4: X1,
               x5: X0, x6: X0, x7: X1, x8: X1};
  x94:Octet = {x1: X1, x2: X0, x3: X0, x4: X1,
               x5: X0, x6: X1, x7: X0, x8: X0};
  x95:Octet = {x1: X1, x2: X0, x3: X0, x4: X1,
               x5: X0, x6: X1, x7: X0, x8: X1};
  x96:Octet = {x1: X1, x2: X0, x3: X0, x4: X1,
               x5: X0, x6: X1, x7: X1, x8: X0};
  x97:Octet = {x1: X1, x2: X0, x3: X0, x4: X1,
               x5: X0, x6: X1, x7: X1, x8: X1};
  x98:Octet = {x1: X1, x2: X0, x3: X0, x4: X1,
               x5: X1, x6: X0, x7: X0, x8: X0};
  x99:Octet = {x1: X1, x2: X0, x3: X0, x4: X1,
               x5: X1, x6: X0, x7: X0, x8: X1};
  x9A:Octet = {x1: X1, x2: X0, x3: X0, x4: X1,
               x5: X1, x6: X0, x7: X1, x8: X0};
  x9B:Octet = {x1: X1, x2: X0, x3: X0, x4: X1,
               x5: X1, x6: X0, x7: X1, x8: X1};
  x9C:Octet = {x1: X1, x2: X0, x3: X0, x4: X1,
               x5: X1, x6: X1, x7: X0, x8: X0};
  x9D:Octet = {x1: X1, x2: X0, x3: X0, x4: X1,
               x5: X1, x6: X1, x7: X0, x8: X1};
  x9E:Octet = {x1: X1, x2: X0, x3: X0, x4: X1,
               x5: X1, x6: X1, x7: X1, x8: X0};
  x9F:Octet = {x1: X1, x2: X0, x3: X0, x4: X1,
               x5: X1, x6: X1, x7: X1, x8: X1};
  xA1:Octet = {x1: X1, x2: X0, x3: X1, x4: X0,
               x5: X0, x6: X0, x7: X0, x8: X1};
  xA0:Octet = {x1: X1, x2: X0, x3: X1, x4: X0,
               x5: X0, x6: X0, x7: X0, x8: X0};
  xA3:Octet = {x1: X1, x2: X0, x3: X1, x4: X0,
               x5: X0, x6: X0, x7: X1, x8: X1};
  xA4:Octet = {x1: X1, x2: X0, x3: X1, x4: X0,
               x5: X0, x6: X1, x7: X0, x8: X0};
  xA5:Octet = {x1: X1, x2: X0, x3: X1, x4: X0,
               x5: X0, x6: X1, x7: X0, x8: X1};
  xA6:Octet = {x1: X1, x2: X0, x3: X1, x4: X0,
               x5: X0, x6: X1, x7: X1, x8: X0};
  xA7:Octet = {x1: X1, x2: X0, x3: X1, x4: X0,
               x5: X0, x6: X1, x7: X1, x8: X1};
  xA8:Octet = {x1: X1, x2: X0, x3: X1, x4: X0,
               x5: X1, x6: X0, x7: X0, x8: X0};
  xA9:Octet = {x1: X1, x2: X0, x3: X1, x4: X0,
               x5: X1, x6: X0, x7: X0, x8: X1};
  xAA:Octet = {x1: X1, x2: X0, x3: X1, x4: X0,
               x5: X1, x6: X0, x7: X1, x8: X0};
  xAB:Octet = {x1: X1, x2: X0, x3: X1, x4: X0,
               x5: X1, x6: X0, x7: X1, x8: X1};
  xAC:Octet = {x1: X1, x2: X0, x3: X1, x4: X0,
               x5: X1, x6: X1, x7: X0, x8: X0};
  xAE:Octet = {x1: X1, x2: X0, x3: X1, x4: X0,
               x5: X1, x6: X1, x7: X1, x8: X0};
  xAF:Octet = {x1: X1, x2: X0, x3: X1, x4: X0,
               x5: X1, x6: X1, x7: X1, x8: X1};
  xB0:Octet = {x1: X1, x2: X0, x3: X1, x4: X1,
               x5: X0, x6: X0, x7: X0, x8: X0};
  xB1:Octet = {x1: X1, x2: X0, x3: X1, x4: X1,
               x5: X0, x6: X0, x7: X0, x8: X1};
  xB2:Octet = {x1: X1, x2: X0, x3: X1, x4: X1,
               x5: X0, x6: X0, x7: X1, x8: X0};
  xB3:Octet = {x1: X1, x2: X0, x3: X1, x4: X1,
               x5: X0, x6: X0, x7: X1, x8: X1};
  xB5:Octet = {x1: X1, x2: X0, x3: X1, x4: X1,
               x5: X0, x6: X1, x7: X0, x8: X1};
  xB6:Octet = {x1: X1, x2: X0, x3: X1, x4: X1,
               x5: X0, x6: X1, x7: X1, x8: X0};
  xB8:Octet = {x1: X1, x2: X0, x3: X1, x4: X1,
               x5: X1, x6: X0, x7: X0, x8: X0};
  xB9:Octet = {x1: X1, x2: X0, x3: X1, x4: X1,
               x5: X1, x6: X0, x7: X0, x8: X1};
  xBA:Octet = {x1: X1, x2: X0, x3: X1, x4: X1,
               x5: X1, x6: X0, x7: X1, x8: X0};
  xBB:Octet = {x1: X1, x2: X0, x3: X1, x4: X1,
               x5: X1, x6: X0, x7: X1, x8: X1};
  xBC:Octet = {x1: X1, x2: X0, x3: X1, x4: X1,
               x5: X1, x6: X1, x7: X0, x8: X0};
  xBE:Octet = {x1: X1, x2: X0, x3: X1, x4: X1,
               x5: X1, x6: X1, x7: X1, x8: X0};
  xBF:Octet = {x1: X1, x2: X0, x3: X1, x4: X1,
               x5: X1, x6: X1, x7: X1, x8: X1};
  xC0:Octet = {x1: X1, x2: X1, x3: X0, x4: X0,
               x5: X0, x6: X0, x7: X0, x8: X0};
  xC1:Octet = {x1: X1, x2: X1, x3: X0, x4: X0,
               x5: X0, x6: X0, x7: X0, x8: X1};
  xC2:Octet = {x1: X1, x2: X1, x3: X0, x4: X0,
               x5: X0, x6: X0, x7: X1, x8: X0};
  xC4:Octet = {x1: X1, x2: X1, x3: X0, x4: X0,
               x5: X0, x6: X1, x7: X0, x8: X0};
  xC5:Octet = {x1: X1, x2: X1, x3: X0, x4: X0,
               x5: X0, x6: X1, x7: X0, x8: X1};
  xC6:Octet = {x1: X1, x2: X1, x3: X0, x4: X0,
               x5: X0, x6: X1, x7: X1, x8: X0};
  xC7:Octet = {x1: X1, x2: X1, x3: X0, x4: X0,
               x5: X0, x6: X1, x7: X1, x8: X1};
  xC8:Octet = {x1: X1, x2: X1, x3: X0, x4: X0,
               x5: X1, x6: X0, x7: X0, x8: X0};
  xC9:Octet = {x1: X1, x2: X1, x3: X0, x4: X0,
               x5: X1, x6: X0, x7: X0, x8: X1};
  xCA:Octet = {x1: X1, x2: X1, x3: X0, x4: X0,
               x5: X1, x6: X0, x7: X1, x8: X0};
  xCB:Octet = {x1: X1, x2: X1, x3: X0, x4: X0,
               x5: X1, x6: X0, x7: X1, x8: X1};
  xCC:Octet = {x1: X1, x2: X1, x3: X0, x4: X0,
               x5: X1, x6: X1, x7: X0, x8: X0};
  xCD:Octet = {x1: X1, x2: X1, x3: X0, x4: X0,
               x5: X1, x6: X1, x7: X0, x8: X1};
  xCE:Octet = {x1: X1, x2: X1, x3: X0, x4: X0,
               x5: X1, x6: X1, x7: X1, x8: X0};
  xD0:Octet = {x1: X1, x2: X1, x3: X0, x4: X1,
               x5: X0, x6: X0, x7: X0, x8: X0};
  xD1:Octet = {x1: X1, x2: X1, x3: X0, x4: X1,
               x5: X0, x6: X0, x7: X0, x8: X1};
  xD2:Octet = {x1: X1, x2: X1, x3: X0, x4: X1,
               x5: X0, x6: X0, x7: X1, x8: X0};
  xD3:Octet = {x1: X1, x2: X1, x3: X0, x4: X1,
               x5: X0, x6: X0, x7: X1, x8: X1};
  xD4:Octet = {x1: X1, x2: X1, x3: X0, x4: X1,
               x5: X0, x6: X1, x7: X0, x8: X0};
  xD5:Octet = {x1: X1, x2: X1, x3: X0, x4: X1,
               x5: X0, x6: X1, x7: X0, x8: X1};
  xD6:Octet = {x1: X1, x2: X1, x3: X0, x4: X1,
               x5: X0, x6: X1, x7: X1, x8: X0};
  xD7:Octet = {x1: X1, x2: X1, x3: X0, x4: X1,
               x5: X0, x6: X1, x7: X1, x8: X1};
  xD9:Octet = {x1: X1, x2: X1, x3: X0, x4: X1,
               x5: X1, x6: X0, x7: X0, x8: X1};
  xD8:Octet = {x1: X1, x2: X1, x3: X0, x4: X1,
               x5: X1, x6: X0, x7: X0, x8: X0};
  xDB:Octet = {x1: X1, x2: X1, x3: X0, x4: X1,
               x5: X1, x6: X0, x7: X1, x8: X1};
  xDC:Octet = {x1: X1, x2: X1, x3: X0, x4: X1,
               x5: X1, x6: X1, x7: X0, x8: X0};
  xDD:Octet = {x1: X1, x2: X1, x3: X0, x4: X1,
               x5: X1, x6: X1, x7: X0, x8: X1};
  xDE:Octet = {x1: X1, x2: X1, x3: X0, x4: X1,
               x5: X1, x6: X1, x7: X1, x8: X0};
  xDF:Octet = {x1: X1, x2: X1, x3: X0, x4: X1,
               x5: X1, x6: X1, x7: X1, x8: X1};
  xE0:Octet = {x1: X1, x2: X1, x3: X1, x4: X0,
               x5: X0, x6: X0, x7: X0, x8: X0};
  xE1:Octet = {x1: X1, x2: X1, x3: X1, x4: X0,
               x5: X0, x6: X0, x7: X0, x8: X1};
  xE2:Octet = {x1: X1, x2: X1, x3: X1, x4: X0,
               x5: X0, x6: X0, x7: X1, x8: X0};
  xE3:Octet = {x1: X1, x2: X1, x3: X1, x4: X0,
               x5: X0, x6: X0, x7: X1, x8: X1};
  xE6:Octet = {x1: X1, x2: X1, x3: X1, x4: X0,
               x5: X0, x6: X1, x7: X1, x8: X0};
  xE8:Octet = {x1: X1, x2: X1, x3: X1, x4: X0,
               x5: X1, x6: X0, x7: X0, x8: X0};
  xE9:Octet = {x1: X1, x2: X1, x3: X1, x4: X0,
               x5: X1, x6: X0, x7: X0, x8: X1};
  xEA:Octet = {x1: X1, x2: X1, x3: X1, x4: X0,
               x5: X1, x6: X0, x7: X1, x8: X0};
  xEB:Octet = {x1: X1, x2: X1, x3: X1, x4: X0,
               x5: X1, x6: X0, x7: X1, x8: X1};
  xEC:Octet = {x1: X1, x2: X1, x3: X1, x4: X0,
               x5: X1, x6: X1, x7: X0, x8: X0};
  xED:Octet = {x1: X1, x2: X1, x3: X1, x4: X0,
               x5: X1, x6: X1, x7: X0, x8: X1};
  xEE:Octet = {x1: X1, x2: X1, x3: X1, x4: X0,
               x5: X1, x6: X1, x7: X1, x8: X0};
  xEF:Octet = {x1: X1, x2: X1, x3: X1, x4: X0,
               x5: X1, x6: X1, x7: X1, x8: X1};
  xF0:Octet = {x1: X1, x2: X1, x3: X1, x4: X1,
               x5: X0, x6: X0, x7: X0, x8: X0};
  xF1:Octet = {x1: X1, x2: X1, x3: X1, x4: X1,
               x5: X0, x6: X0, x7: X0, x8: X1};
  xF2:Octet = {x1: X1, x2: X1, x3: X1, x4: X1,
               x5: X0, x6: X0, x7: X1, x8: X0};
  xF3:Octet = {x1: X1, x2: X1, x3: X1, x4: X1,
               x5: X0, x6: X0, x7: X1, x8: X1};
  xF4:Octet = {x1: X1, x2: X1, x3: X1, x4: X1,
               x5: X0, x6: X1, x7: X0, x8: X0};
  xF5:Octet = {x1: X1, x2: X1, x3: X1, x4: X1,
               x5: X0, x6: X1, x7: X0, x8: X1};
  xF6:Octet = {x1: X1, x2: X1, x3: X1, x4: X1,
               x5: X0, x6: X1, x7: X1, x8: X0};
  xF7:Octet = {x1: X1, x2: X1, x3: X1, x4: X1,
               x5: X0, x6: X1, x7: X1, x8: X1};
  xF8:Octet = {x1: X1, x2: X1, x3: X1, x4: X1,
               x5: X1, x6: X0, x7: X0, x8: X0};
  xF9:Octet = {x1: X1, x2: X1, x3: X1, x4: X1,
               x5: X1, x6: X0, x7: X0, x8: X1};
  xFA:Octet = {x1: X1, x2: X1, x3: X1, x4: X1,
               x5: X1, x6: X0, x7: X1, x8: X0};
  xFB:Octet = {x1: X1, x2: X1, x3: X1, x4: X1,
               x5: X1, x6: X0, x7: X1, x8: X1};
  xFC:Octet = {x1: X1, x2: X1, x3: X1, x4: X1,
               x5: X1, x6: X1, x7: X0, x8: X0};
  xFD:Octet = {x1: X1, x2: X1, x3: X1, x4: X1,
               x5: X1, x6: X1, x7: X0, x8: X1};
  xFE:Octet = {x1: X1, x2: X1, x3: X1, x4: X1,
               x5: X1, x6: X1, x7: X1, x8: X0};
  xFF:Octet = {x1: X1, x2: X1, x3: X1, x4: X1,
               x5: X1, x6: X1, x7: X1, x8: X1};
------------------------------------------
function eqOctet (o1, o2: Octet) returns (res: bool)
  res =  ((o1.x1 = o2.x1) and (o1.x2 = o2.x2) and (o1.x3 = o2.x3) and
         (o1.x4 = o2.x4) and (o1.x5 = o2.x5) and (o1.x6 = o2.x6)
         and (o1.x7 = o2.x7) and (o1.x8 = o2.x8));
------------------------------------------
function andOctet (o1, o2: Octet) returns (o: Octet)
  o = {x1: andBit (o1.x1, o2.x1), x2: andBit (o1.x2, o2.x2),
       x3: andBit (o1.x3, o2.x3), x4: andBit (o1.x4, o2.x4),
       x5: andBit (o1.x5, o2.x5), x6: andBit (o1.x6, o2.x6),
       x7: andBit (o1.x7, o2.x7), x8: andBit (o1.x8, o2.x8)};
------------------------------------------
function orOctet (o1, o2: Octet) returns (o: Octet)
  o = {x1: orBit (o1.x1, o2.x1), x2: orBit (o1.x2, o2.x2),
       x3: orBit (o1.x3, o2.x3), x4: orBit (o1.x4, o2.x4),
       x5: orBit (o1.x5, o2.x5), x6: orBit (o1.x6, o2.x6),
       x7: orBit (o1.x7, o2.x7), x8: orBit (o1.x8, o2.x8)};

------------------------------------------
function xorOctet (o1, o2: Octet) returns (o: Octet)
  o = {x1: xorBit (o1.x1, o2.x1), x2: xorBit (o1.x2, o2.x2),
       x3: xorBit (o1.x3, o2.x3), x4: xorBit (o1.x4, o2.x4),
       x5: xorBit (o1.x5, o2.x5), x6: xorBit (o1.x6, o2.x6),
       x7: xorBit (o1.x7, o2.x7), x8: xorBit (o1.x8, o2.x8)};
------------------------------------------
function leftOctet1 (o1: Octet) returns (o: Octet)
  o  = {x1: o1.x2, x2: o1.x3, x3: o1.x4, x4: o1.x5,
        x5: o1.x6, x6: o1.x7, x7: o1.x8, x8: X0};
------------------------------------------
function leftOctet2 (o1: Octet) returns (o: Octet)
  o  = {x1: o1.x3, x2: o1.x4, x3: o1.x5, x4: o1.x6,
        x5: o1.x7, x6: o1.x8, x7: X0, x8: X0};
------------------------------------------
function leftOctet3 (o1: Octet) returns (o: Octet)
  o  = {x1: o1.x4, x2: o1.x5, x3: o1.x6, x4: o1.x7,
        x5: o1.x8, x6: X0, x7: X0, x8: X0};
------------------------------------------
function leftOctet4 (o1: Octet) returns (o: Octet)
  o  = {x1: o1.x5, x2: o1.x6, x3: o1.x7, x4: o1.x8,
        x5: X0, x6: X0, x7: X0, x8: X0};
------------------------------------------
function leftOctet5 (o1: Octet) returns (o: Octet)
  o  = {x1: o1.x6, x2: o1.x7, x3: o1.x8, x4: X0,
        x5: X0, x6: X0, x7: X0, x8: X0};
------------------------------------------
function leftOctet6 (o1: Octet) returns (o: Octet)
  o  = {x1: o1.x7, x2: o1.x8, x3: X0, x4: X0,
        x5: X0, x6: X0, x7: X0, x8: X0};
------------------------------------------
function leftOctet7 (o1: Octet) returns (o: Octet)
  o  = {x1: o1.x8, x2: X0, x3: X0, x4: X0,
        x5: X0, x6: X0, x7: X0, x8: X0};
------------------------------------------
function rightOctet1 (o1: Octet) returns (o: Octet)
  o  = {x1: X0, x2: o1.x1, x3: o1.x2, x4: o1.x3,
        x5: o1.x4, x6: o1.x5, x7: o1.x6, x8: o1.x7};
------------------------------------------
function rightOctet2 (o1: Octet) returns (o: Octet)
  o  = {x1: X0, x2: X0, x3: o1.x1, x4: o1.x2,
        x5: o1.x3, x6: o1.x4, x7: o1.x5, x8: o1.x6};
------------------------------------------
function rightOctet3 (o1: Octet) returns (o: Octet)
  o  = {x1: X0, x2: X0, x3: X0, x4: o1.x1,
        x5: o1.x2, x6: o1.x3, x7: o1.x4, x8: o1.x5};
------------------------------------------
function rightOctet4 (o1: Octet) returns (o: Octet)
  o  = {x1: X0, x2: X0, x3: X0, x4: X0,
        x5: o1.x1, x6: o1.x2, x7: o1.x3, x8: o1.x4};
------------------------------------------
function rightOctet5 (o1: Octet) returns (o: Octet)
  o  = {x1: X0, x2: X0, x3: X0, x4: X0,
        x5: X0, x6: o1.x1, x7: o1.x2, x8: o1.x3};
------------------------------------------
function rightOctet6 (o1: Octet) returns (o: Octet)
  o  = {x1: X0, x2: X0, x3: X0, x4: X0,
        x5: X0, x6: X0, x7: o1.x1, x8: o1.x2};
------------------------------------------
function rightOctet7 (o1: Octet) returns (o: Octet)
  o  = {x1: X0, x2: X0, x3: X0, x4: X0,
        x5: X0, x6: X0, x7: X0, x8: o1.x1};
\end{lstlisting}

\section{Definitions for Type OctetSum}
We define type OctetSum that stores the result of the addition of two octets. 
Values of this type are 9-bit words, made up using the structure \lstinline+OctetSum+ that gathers one bit for the carry and an octet for the sum.
The two principal functions for this type are addOctetSum (which adds two octets and an input carry bit, and returns both an output carry bit and an 8-bit sum), and addOctet (which is derived from the former one by dropping the input and output carry bits); the other functions are auxiliary functions implementing an 8-bit adder.
\begin{lstlisting}[language=Lustre]
------------------------------------------
type OctetSum = {x: Bit, o: Octet};
------------------------------------------
function addBit (x1, x2, x3: Bit) returns (x: Bit)
  x = xorBit (xorBit (x1, x2), x3);
------------------------------------------
function carBit (x1, x2, x3: Bit) returns (x: Bit)
  x = orBit (andBit (andBit (x1, x2), notBit (x3)),
             andBit (orBit (x1, x2), x3));
------------------------------------------
function addOctetSum (o1, o2: Octet; x: Bit) returns (os: OctetSum)
var x1, x11, x2, x22, x3, x33, x4, x44, x5, x55: Bit;
    x6, x66, x7, x77, x8, x88: Bit;
let
  x1 = carBit (o1.x8, o2.x8, x);
  x11 = addBit (o1.x8, o2.x8, x);
  x2 =  carBit (o1.x7, o2.x7, x1);
  x22 = addBit (o1.x7, o2.x7, x1);
  x3 =  carBit (o1.x6, o2.x6, x2);
  x33 = addBit (o1.x6, o2.x6, x2);
  x4 =  carBit (o1.x5, o2.x5, x3);
  x44 = addBit (o1.x5, o2.x5, x3);
  x5 =  carBit (o1.x4, o2.x4, x4);
  x55 = addBit (o1.x4, o2.x4, x4);
  x6 =  carBit (o1.x3, o2.x3, x5);
  x66 = addBit (o1.x3, o2.x3, x5);
  x7 =  carBit (o1.x2, o2.x2, x6);
  x77 = addBit (o1.x2, o2.x2, x6);
  x8 =  carBit (o1.x1, o2.x1, x7);
  x88 = addBit (o1.x1, o2.x1, x7);
  os = {x: x8, o: {x1: x88, x2: x77, x3: x66, x4: x55,
                   x5: x44, x6: x33, x7: x22, x8: x11}};
tel;
------------------------------------------
function dropCarryOctetSum (os: OctetSum) returns (o: Octet)
  o = os.o;
------------------------------------------
function addOctet (o1, o2: Octet) returns (o: Octet)
  o = dropCarryOctetSum (addOctetSum (o1, o2, X0));
------------------------------------------
\end{lstlisting}

\section{Definitions for Type Half}
\label{sec:16bitmul}
We define 16-bit words (``named half words'') using a structure \lstinline+Half+ that contains two bytes corresponding to a half word, together with two usual constants, and a function implementing operation mulOctet that takes two octets and computes their 16-bit product; the other functions are auxiliary functions implementing an 8-bit multiplier.
\begin{lstlisting}[language=Lustre]
------------------------------------------
type Half = {o1: Octet, o2: Octet};
------------------------------------------
const 
  x0000:Half = {o1: x00, o2: x00};
  x0001:Half = {o1: x00, o2: x01};
------------------------------------------
function mulOctetA (h1: Half; o1, o2: Octet) returns (h: Half)
var o3: Octet; os: OctetSum;
let
  o3 = addOctet (h1.o1, o1);
  os = addOctetSum (h1.o2, o2, X0);
  h = if os.x = X0 then {o1: o3, o2: os.o} 
      else {o1: addOctet (o3, x01), o2: os.o};
tel;
------------------------------------------
function mulOctet (o1, o2: Octet) returns (h: Half)
var h1, h2, h3, h4, h5, h6, h7: Half;
let
  h1 = if o1.x1 = X0 then x0000 
       else  mulOctetA (x0000, rightOctet1 (o2), leftOctet7 (o2));
  h2 = if o1.x2 = X0 then h1 
       else  mulOctetA (h1, rightOctet2 (o2), leftOctet6 (o2));
  h3 = if o1.x3 = X0 then h2 
       else  mulOctetA (h2, rightOctet3 (o2), leftOctet5 (o2));
  h4 = if o1.x4 = X0 then h3 
       else  mulOctetA (h3, rightOctet4 (o2), leftOctet4 (o2));
  h5 = if o1.x5 = X0 then h4 
       else  mulOctetA (h4, rightOctet5 (o2), leftOctet3 (o2));
  h6 = if o1.x6 = X0 then h5 
       else  mulOctetA (h5, rightOctet6 (o2), leftOctet2 (o2));
  h7 = if o1.x7 = X0 then h6 
       else  mulOctetA (h6, rightOctet7 (o2), leftOctet1 (o2));
  h = if o1.x8 = X0 then h7 
       else  mulOctetA (h7, x00, o2);
tel;
------------------------------------------
\end{lstlisting}

\section{Definitions for Type HalfSum}
We define type HalfSum that stores the result of the addition of two half words. 
Values of this type are 17-bit words, made up using the constructor buildHalfSum that gathers one bit for the carry and a half word for the sum.
The five principal non-constructors for this type are eqHalfSum (which tests equality), addHalfSum (which adds two half words and returns both a carry bit and a 16-bit sum), addHalf (which is derived from the former one by dropping the carry bit), addHalfOctet and addHalfOctets (which are similar to the former one but take byte arguments that are converted to half words before summation); the other non-constructors are auxiliary functions implementing a 16-bit adder built using two 8-bit adders.
\begin{lstlisting}[language=Lustre]
------------------------------------------
type HalfSum = {x: Bit, h: Half};
------------------------------------------
function addHalfSum (h1, h2: Half) returns (hs: HalfSum)
var os, os1: OctetSum;
let
  os = addOctetSum (h1.o2, h2.o2, X0);
  os1 = addOctetSum (h1.o1, h2.o1, os.x);
  hs = {x: os1.x, h: {o1: os1.o, o2: os.o}};
tel;
------------------------------------------
function dropCarryHalfSum (hs: HalfSum) returns (h: Half)
  h = hs.h;
------------------------------------------
function addHalf (h1, h2: Half) returns (h: Half)
  h = dropCarryHalfSum (addHalfSum (h1, h2));
------------------------------------------
function addHalfOctet (o1: Octet; h1: Half) returns (h: Half)
  h = addHalf ({o1: x00, o2: o1},  h1);
------------------------------------------
function addHalfOctets (o1, o2: Octet) returns (h: Half)
  h = addHalf ({o1: x00, o2: o1}, {o1: x00, o2: o2});
------------------------------------------
\end{lstlisting}
\section{Definitions for Type Block}
\label{sec:32bitmul}
We define 32-bit words (named “blocks” according to the MAA terminology) using a constructor buildBlock that takes four bytes and returns a block.
The seven principal non-constructors for this type are eqBlock (which tests equality), andBlock, orBlock, and xorBlock (which implement bitwise logical operations on blocks), HalfU and HalfL (which decompose a block into two half words), and mulHalf (which takes two half words and computes their 32-bit product); the other non-constructors are auxiliary functions implementing a 16-bit multiplier built using four 8-bit multipliers, as well as all block constants needed to formally describe the MAA and its test vectors.
\begin{lstlisting}[language=Lustre]
------------------------------------------
type Block = {o1: Octet, o2: Octet, o3: Octet, o4: Octet};
------------------------------------------
const 
  x00000000:Block = {o1: x00, o2: x00, o3: x00, o4: x00};
  x00000001:Block = {o1: x00, o2: x00, o3: x00, o4: x01};
  x00000002:Block = {o1: x00, o2: x00, o3: x00, o4: x02};
  x00000003:Block = {o1: x00, o2: x00, o3: x00, o4: x03};
  x00000004:Block = {o1: x00, o2: x00, o3: x00, o4: x04};
  x00000005:Block = {o1: x00, o2: x00, o3: x00, o4: x05};
  x00000006:Block = {o1: x00, o2: x00, o3: x00, o4: x06};
  x00000007:Block = {o1: x00, o2: x00, o3: x00, o4: x07};
  x00000008:Block = {o1: x00, o2: x00, o3: x00, o4: x08};
  x00000009:Block = {o1: x00, o2: x00, o3: x00, o4: x09};
  x0000000A:Block = {o1: x00, o2: x00, o3: x00, o4: x0A};
  x0000000B:Block = {o1: x00, o2: x00, o3: x00, o4: x0B};
  x0000000C:Block = {o1: x00, o2: x00, o3: x00, o4: x0C};
  x0000000D:Block = {o1: x00, o2: x00, o3: x00, o4: x0D};
  x0000000E:Block = {o1: x00, o2: x00, o3: x00, o4: x0E};
  x0000000F:Block = {o1: x00, o2: x00, o3: x00, o4: x0F};
  x00000010:Block = {o1: x00, o2: x00, o3: x00, o4: x10};
  x00000012:Block = {o1: x00, o2: x00, o3: x00, o4: x12};
  x00000014:Block = {o1: x00, o2: x00, o3: x00, o4: x14};
  x00000016:Block = {o1: x00, o2: x00, o3: x00, o4: x16};
  x00000018:Block = {o1: x00, o2: x00, o3: x00, o4: x18};
  x0000001B:Block = {o1: x00, o2: x00, o3: x00, o4: x1B};
  x0000001D:Block = {o1: x00, o2: x00, o3: x00, o4: x1D};
  x0000001E:Block = {o1: x00, o2: x00, o3: x00, o4: x1E};
  x0000001F:Block = {o1: x00, o2: x00, o3: x00, o4: x1F};
  x00000031:Block = {o1: x00, o2: x00, o3: x00, o4: x31};
  x00000036:Block = {o1: x00, o2: x00, o3: x00, o4: x36};
  x00000060:Block = {o1: x00, o2: x00, o3: x00, o4: x60};
  x00000080:Block = {o1: x00, o2: x00, o3: x00, o4: x80};
  x000000A5:Block = {o1: x00, o2: x00, o3: x00, o4: xA5};
  x000000B6:Block = {o1: x00, o2: x00, o3: x00, o4: xB6};
  x000000C4:Block = {o1: x00, o2: x00, o3: x00, o4: xC4};
  x000000D2:Block = {o1: x00, o2: x00, o3: x00, o4: xD2};
  x00000100:Block = {o1: x00, o2: x00, o3: x01, o4: x00};
  x00000129:Block = {o1: x00, o2: x00, o3: x01, o4: x29};
  x0000018C:Block = {o1: x00, o2: x00, o3: x01, o4: x8C};
  x00004000:Block = {o1: x00, o2: x00, o3: x40, o4: x00};
  x00010000:Block = {o1: x00, o2: x01, o3: x00, o4: x00};
  x00020000:Block = {o1: x00, o2: x02, o3: x00, o4: x00};
  x00030000:Block = {o1: x00, o2: x03, o3: x00, o4: x00};
  x00040000:Block = {o1: x00, o2: x04, o3: x00, o4: x00};
  x00060000:Block = {o1: x00, o2: x06, o3: x00, o4: x00};
  x00804021:Block = {o1: x00, o2: x80, o3: x40, o4: x21};
  x00FF00FF:Block = {o1: x00, o2: xFF, o3: x00, o4: xFF};
  x0103050B:Block = {o1: x01, o2: x03, o3: x05, o4: x0B};
  x01030703:Block = {o1: x01, o2: x03, o3: x07, o4: x03};
  x01030705:Block = {o1: x01, o2: x03, o3: x07, o4: x05};
  x0103070F:Block = {o1: x01, o2: x03, o3: x07, o4: x0F};
  x02040801:Block = {o1: x02, o2: x04, o3: x08, o4: x01};
  x0297AF6F:Block = {o1: x02, o2: x97, o3: xAF, o4: x6F};
  x07050301:Block = {o1: x07, o2: x05, o3: x03, o4: x01};
  x07C72EAA:Block = {o1: x07, o2: xC7, o3: x2E, o4: xAA};
  x0A202020:Block = {o1: x0A, o2: x20, o3: x20, o4: x20};
  x0AD67E20:Block = {o1: x0A, o2: xD6, o3: x7E, o4: x20};
  x10000000:Block = {o1: x10, o2: x00, o3: x00, o4: x00};
  x11A9D254:Block = {o1: x11, o2: xA9, o3: xD2, o4: x54};
  x11AC46B8:Block = {o1: x11, o2: xAC, o3: x46, o4: xB8};
  x1277A6D4:Block = {o1: x12, o2: x77, o3: xA6, o4: xD4};
  x13647149:Block = {o1: x13, o2: x64, o3: x71, o4: x49};
  x160EE9B5:Block = {o1: x16, o2: x0E, o3: xE9, o4: xB5};
  x17065DBB:Block = {o1: x17, o2: x06, o3: x5D, o4: xBB};
  x1D10D8D3:Block = {o1: x1D, o2: x10, o3: xD8, o4: xD3};
  x1D3B7760:Block = {o1: x1D, o2: x3B, o3: x77, o4: x60};
  x1D9C9655:Block = {o1: x1D, o2: x9C, o3: x96, o4: x55};
  x1F3F7FFF:Block = {o1: x1F, o2: x3F, o3: x7F, o4: xFF};
  x21D869BA:Block = {o1: x21, o2: xD8, o3: x69, o4: xBA};
  x24B66FB5:Block = {o1: x24, o2: xB6, o3: x6F, o4: xB5};
  x270EEDAF:Block = {o1: x27, o2: x0E, o3: xED, o4: xAF};
  x277B4B25:Block = {o1: x27, o2: x7B, o3: x4B, o4: x25};
  x2829040B:Block = {o1: x28, o2: x29, o3: x04, o4: x0B};
  x288FC786:Block = {o1: x28, o2: x8F, o3: xC7, o4: x86};
  x28EAD8B3:Block = {o1: x28, o2: xEA, o3: xD8, o4: xB3};
  x29907CD8:Block = {o1: x29, o2: x90, o3: x7C, o4: xD8};
  x29C1485F:Block = {o1: x29, o2: xC1, o3: x48, o4: x5F};
  x29EEE96B:Block = {o1: x29, o2: xEE, o3: xE9, o4: x6B};
  x2A6091AE:Block = {o1: x2A, o2: x60, o3: x91, o4: xAE};
  x2BF8499A:Block = {o1: x2B, o2: xF8, o3: x49, o4: x9A};
  x2E80AC30:Block = {o1: x2E, o2: x80, o3: xAC, o4: x30};
  x2FD76FFB:Block = {o1: x2F, o2: xD7, o3: x6F, o4: xFB};
  x30261492:Block = {o1: x30, o2: x26, o3: x14, o4: x92};
  x303FF4AA:Block = {o1: x30, o2: x3F, o3: xF4, o4: xAA};
  x33D5A466:Block = {o1: x33, o2: xD5, o3: xA4, o4: x66};
  x344925FC:Block = {o1: x34, o2: x49, o3: x25, o4: xFC};
  x34ACF886:Block = {o1: x34, o2: xAC, o3: xF8, o4: x86};
  x3CD54DEB:Block = {o1: x3C, o2: xD5, o3: x4D, o4: xEB};
  x3CF3A7D2:Block = {o1: x3C, o2: xF3, o3: xA7, o4: xD2};
  x3DD81AC6:Block = {o1: x3D, o2: xD8, o3: x1A, o4: xC6};
  x3F6F7248:Block = {o1: x3F, o2: x6F, o3: x72, o4: x48};
  x48B204D6:Block = {o1: x48, o2: xB2, o3: x04, o4: xD6};
  x4A645A01:Block = {o1: x4A, o2: x64, o3: x5A, o4: x01};
  x4C49AAE0:Block = {o1: x4C, o2: x49, o3: xAA, o4: xE0};
  x4CE933E1:Block = {o1: x4C, o2: xE9, o3: x33, o4: xE1};
  x4D53901A:Block = {o1: x4D, o2: x53, o3: x90, o4: x1A};
  x4DA124A1:Block = {o1: x4D, o2: xA1, o3: x24, o4: xA1};
  x4F998E01:Block = {o1: x4F, o2: x99, o3: x8E, o4: x01};
  x50DEC930:Block = {o1: x50, o2: xDE, o3: xC9, o4: x30};
  x51AF3C1D:Block = {o1: x51, o2: xAF, o3: x3C, o4: x1D};
  x51EDE9C7:Block = {o1: x51, o2: xED, o3: xE9, o4: xC7};
  x550D91CE:Block = {o1: x55, o2: x0D, o3: x91, o4: xCE};
  x55555555:Block = {o1: x55, o2: x55, o3: x55, o4: x55};
  x55DD063F:Block = {o1: x55, o2: xDD, o3: x06, o4: x3F};
  x5834A585:Block = {o1: x58, o2: x34, o3: xA5, o4: x85};
  x5A35D667:Block = {o1: x5A, o2: x35, o3: xD6, o4: x67};
  x5BC02502:Block = {o1: x5B, o2: xC0, o3: x25, o4: x02};
  x5CCA3239:Block = {o1: x5C, o2: xCA, o3: x32, o4: x39};
  x5EBA06C2:Block = {o1: x5E, o2: xBA, o3: x06, o4: xC2};
  xF0239DD5:Block = {o1: xF0, o2: x23, o3: x9D, o4: xD5};
  x5F38EEF1:Block = {o1: x5F, o2: x38, o3: xEE, o4: xF1};
  x613F8E2A:Block = {o1: x61, o2: x3F, o3: x8E, o4: x2A};
  x63C70DBA:Block = {o1: x63, o2: xC7, o3: x0D, o4: xBA};
  x6AD6E8A4:Block = {o1: x6A, o2: xD6, o3: xE8, o4: xA4};
  x6AEBACF8:Block = {o1: x6A, o2: xEB, o3: xAC, o4: xF8};
  x6D67E884:Block = {o1: x6D, o2: x67, o3: xE8, o4: x84};
  x7050EC5E:Block = {o1: x70, o2: x50, o3: xEC, o4: x5E};
  x717153D5:Block = {o1: x71, o2: x71, o3: x53, o4: xD5};
  x7201F4DC:Block = {o1: x72, o2: x01, o3: xF4, o4: xDC};
  x7397C9AE:Block = {o1: x73, o2: x97, o3: xC9, o4: xAE};
  x74B39176:Block = {o1: x74, o2: xB3, o3: x91, o4: x76};
  x7783C51D:Block = {o1: x77, o2: x83, o3: xC5, o4: x1D};
  x7792F9D4:Block = {o1: x77, o2: x92, o3: xF9, o4: xD4};
  x7BC180AB:Block = {o1: x7B, o2: xC1, o3: x80, o4: xAB};
  x7DB2D9F4:Block = {o1: x7D, o2: xB2, o3: xD9, o4: xF4};
  x7DFEFBFF:Block = {o1: x7D, o2: xFE, o3: xFB, o4: xFF};
  x7F76A3B0:Block = {o1: x7F, o2: x76, o3: xA3, o4: xB0};
  x7F839576:Block = {o1: x7F, o2: x83, o3: x95, o4: x76};
  x7FFFFFF0:Block = {o1: x7F, o2: xFF, o3: xFF, o4: xF0};
  x7FFFFFF1:Block = {o1: x7F, o2: xFF, o3: xFF, o4: xF1};
  x7FFFFFFC:Block = {o1: x7F, o2: xFF, o3: xFF, o4: xFC};
  x7FFFFFFD:Block = {o1: x7F, o2: xFF, o3: xFF, o4: xFD};
  x80000000:Block = {o1: x80, o2: x00, o3: x00, o4: x00};
  x80000002:Block = {o1: x80, o2: x00, o3: x00, o4: x02};
  x800000C2:Block = {o1: x80, o2: x00, o3: x00, o4: xC2};
  x80018000:Block = {o1: x80, o2: x01, o3: x80, o4: x00};
  x80018001:Block = {o1: x80, o2: x01, o3: x80, o4: x01};
  x80397302:Block = {o1: x80, o2: x39, o3: x73, o4: x02};
  x81D10CA3:Block = {o1: x81, o2: xD1, o3: x0C, o4: xA3};
  x89D635D7:Block = {o1: x89, o2: xD6, o3: x35, o4: xD7};
  x8CE37709:Block = {o1: x8C, o2: xE3, o3: x77, o4: x09};
  x8DC8BBDE:Block = {o1: x8D, o2: xC8, o3: xBB, o4: xDE};
  x9115A558:Block = {o1: x91, o2: x15, o3: xA5, o4: x58};
  x91896CFA:Block = {o1: x91, o2: x89, o3: x6C, o4: xFA};
  x9372CDC6:Block = {o1: x93, o2: x72, o3: xCD, o4: xC6};
  x98D1CC75:Block = {o1: x98, o2: xD1, o3: xCC, o4: x75};
  x9D15C437:Block = {o1: x9D, o2: x15, o3: xC4, o4: x37};
  x9DB15CF6:Block = {o1: x9D, o2: xB1, o3: x5C, o4: xF6};
  x9E2E7B36:Block = {o1: x9E, o2: x2E, o3: x7B, o4: x36};
  xA018C83B:Block = {o1: xA0, o2: x18, o3: xC8, o4: x3B};
  xA0B87B77:Block = {o1: xA0, o2: xB8, o3: x7B, o4: x77};
  xA44AAAC0:Block = {o1: xA4, o2: x4A, o3: xAA, o4: xC0};
  xA511987A:Block = {o1: xA5, o2: x11, o3: x98, o4: x7A};
  xA70FC148:Block = {o1: xA7, o2: x0F, o3: xC1, o4: x48};
  xA93BD410:Block = {o1: xA9, o2: x3B, o3: xD4, o4: x10};
  xAAAAAAAA:Block = {o1: xAA, o2: xAA, o3: xAA, o4: xAA};
  xAB00FFCD:Block = {o1: xAB, o2: x00, o3: xFF, o4: xCD};
  xAB01FCCD:Block = {o1: xAB, o2: x01, o3: xFC, o4: xCD};
  xAB6EED4A:Block = {o1: xAB, o2: x6E, o3: xED, o4: x4A};
  xABEEED6B:Block = {o1: xAB, o2: xEE, o3: xED, o4: x6B};
  xACBC13DD:Block = {o1: xAC, o2: xBC, o3: x13, o4: xDD};
  xB1CC1CC5:Block = {o1: xB1, o2: xCC, o3: x1C, o4: xC5};
  xB8142629:Block = {o1: xB8, o2: x14, o3: x26, o4: x29};
  xB99A62DE:Block = {o1: xB9, o2: x9A, o3: x62, o4: xDE};
  xBA92DB12:Block = {o1: xBA, o2: x92, o3: xDB, o4: x12};
  xBBA57835:Block = {o1: xBB, o2: xA5, o3: x78, o4: x35};
  xBE9F0917:Block = {o1: xBE, o2: x9F, o3: x09, o4: x17};
  xBF2D7D85:Block = {o1: xBF, o2: x2D, o3: x7D, o4: x85};
  xBFEF7FDF:Block = {o1: xBF, o2: xEF, o3: x7F, o4: xDF};
  xC1ED90DD:Block = {o1: xC1, o2: xED, o3: x90, o4: xDD};
  xC21A1846:Block = {o1: xC2, o2: x1A, o3: x18, o4: x46};
  xC4EB1AEB:Block = {o1: xC4, o2: xEB, o3: x1A, o4: xEB};
  xC6B1317E:Block = {o1: xC6, o2: xB1, o3: x31, o4: x7E};
  xCBC865BA:Block = {o1: xCB, o2: xC8, o3: x65, o4: xBA};
  xCD959B46:Block = {o1: xCD, o2: x95, o3: x9B, o4: x46};
  xD0482465:Block = {o1: xD0, o2: x48, o3: x24, o4: x65};
  xD636250D:Block = {o1: xD6, o2: x36, o3: x25, o4: x0D};
  xD7843FDC:Block = {o1: xD7, o2: x84, o3: x3F, o4: xDC};
  xD78634BC:Block = {o1: xD7, o2: x86, o3: x34, o4: xBC};
  xD8804CA5:Block = {o1: xD8, o2: x80, o3: x4C, o4: xA5};
  xDB79FBDC:Block = {o1: xDB, o2: x79, o3: xFB, o4: xDC};
  xDB9102B0:Block = {o1: xDB, o2: x91, o3: x02, o4: xB0};
  xE0C08000:Block = {o1: xE0, o2: xC0, o3: x80, o4: x00};
  xE6A12F07:Block = {o1: xE6, o2: xA1, o3: x2F, o4: x07};
  xEB35B97F:Block = {o1: xEB, o2: x35, o3: xB9, o4: x7F};
  xF14D6E28:Block = {o1: xF1, o2: x4D, o3: x6E, o4: x28};
  xF2EF3501:Block = {o1: xF2, o2: xEF, o3: x35, o4: x01};
  xF6A09667:Block = {o1: xF6, o2: xA0, o3: x96, o4: x67};
  xFD297DA4:Block = {o1: xFD, o2: x29, o3: x7D, o4: xA4};
  xFDC1A8BA:Block = {o1: xFD, o2: xC1, o3: xA8, o4: xBA};
  xFE4E5BDD:Block = {o1: xFE, o2: x4E, o3: x5B, o4: xDD};
  xFECCAA6E:Block = {o1: xFE, o2: xCC, o3: xAA, o4: x6E};
  xFEFC07F0:Block = {o1: xFE, o2: xFC, o3: x07, o4: xF0};
  xFF2D7DA5:Block = {o1: xFF, o2: x2D, o3: x7D, o4: xA5};
  xFFEF0001:Block = {o1: xFF, o2: xEF, o3: x00, o4: x01};
  xFFFF00FF:Block = {o1: xFF, o2: xFF, o3: x00, o4: xFF};
  xFFFFFF2D:Block = {o1: xFF, o2: xFF, o3: xFF, o4: x2D};
  xFFFFFF3A:Block = {o1: xFF, o2: xFF, o3: xFF, o4: x3A};
  xFFFFFFF0:Block = {o1: xFF, o2: xFF, o3: xFF, o4: xF0};
  xFFFFFFF1:Block = {o1: xFF, o2: xFF, o3: xFF, o4: xF1};
  xFFFFFFF4:Block = {o1: xFF, o2: xFF, o3: xFF, o4: xF4};
  xFFFFFFF5:Block = {o1: xFF, o2: xFF, o3: xFF, o4: xF5};
  xFFFFFFF7:Block = {o1: xFF, o2: xFF, o3: xFF, o4: xF7};
  xFFFFFFF9:Block = {o1: xFF, o2: xFF, o3: xFF, o4: xF9};
  xFFFFFFFA:Block = {o1: xFF, o2: xFF, o3: xFF, o4: xFA};
  xFFFFFFFB:Block = {o1: xFF, o2: xFF, o3: xFF, o4: xFB};
  xFFFFFFFC:Block = {o1: xFF, o2: xFF, o3: xFF, o4: xFC};
  xFFFFFFFD:Block = {o1: xFF, o2: xFF, o3: xFF, o4: xFD};
  xFFFFFFFE:Block = {o1: xFF, o2: xFF, o3: xFF, o4: xFE};
  xFFFFFFFF:Block = {o1: xFF, o2: xFF, o3: xFF, o4: xFF};
------------------------------------------
function eqBlock (w1, w2: Block) returns (res: bool)
  res =  ((w1.o1 = w2.o2) and (w1.o2 = w2.o2) and (w1.o3 = w2.o3) and
         (w1.o4 = w2.o4));
------------------------------------------
function andBlock (w1, w2: Block) returns (w: Block)
  w = {o1: andOctet (w1.o1, w2.o1), o2: andOctet (w1.o2, w2.o2),
       o3: andOctet (w1.o3, w2.o3), o4: andOctet (w1.o4, w2.o4)};
------------------------------------------
function orBlock (w1, w2: Block) returns (w: Block)
  w = {o1: orOctet (w1.o1, w2.o1), o2: orOctet (w1.o2, w2.o2),
       o3: orOctet (w1.o3, w2.o3), o4: orOctet (w1.o4, w2.o4)};
------------------------------------------
function xorBlock (w1, w2: Block) returns (w: Block)
  w = {o1: xorOctet (w1.o1, w2.o1), o2: xorOctet (w1.o2, w2.o2),
       o3: xorOctet (w1.o3, w2.o3), o4: xorOctet (w1.o4, w2.o4)};
------------------------------------------
function HalfU (w: Block) returns (o1o2: Half)
  o1o2 = {o1: w.o1, o2: w.o2};
------------------------------------------
function HalfL (w: Block) returns (o3o4: Half)
  o3o4 = {o1: w.o3, o2: w.o4};
------------------------------------------
function mulHalf (h1, h2: Half) returns (w: Block)
var h3, h4, h5, h6, h7, h8, h9: Half;
let
  h3 = mulOctet (h1.o1, h2.o1); 
  h4 = mulOctet (h1.o1, h2.o2);
  h5 = mulOctet (h1.o2, h2.o1);
  h6 = mulOctet (h1.o2, h2.o2);
  h7 = addHalfOctet (h4.o2, addHalfOctets (h5.o2, h6.o1));
  h8 = addHalfOctet (h7.o1, addHalfOctet (h3.o2, 
                              addHalfOctets (h4.o1, h5.o1)));
  h9 = addHalfOctets (h8.o1, h3.o1);
  w = {o1: h9.o2, o2: h8.o2, o3: h7.o2, o4: h6.o2};
tel;
------------------------------------------
\end{lstlisting}

\section{Definitions for Type BlockSum}
We define type BlockSum that stores the result of the addition of two blocks. Values of this type are 33-bit words, made up using the constructor buildBlockSum that gathers one bit for the carry and a block for the sum.
The five principal non-constructors for this type are eqBlockSum (which tests equality), addBlockSum (which adds two blocks and returns both a carry bit and a 32-bit sum), addBlock (which is derived from the former one by dropping the carry bit), addBlockHalf and addBlockHalves (which are similar to the former one but take half-word arguments that are converted to blocks before summation); the other non-constructors are auxiliary functions implementing a 32-bit adder built using four 8-bit adders.
\begin{lstlisting}[language=Lustre]
------------------------------------------
type BlockSum = {x: Bit, w: Block};
------------------------------------------
function addBlockSum (w1, w2 : Block) returns (ws: BlockSum)
var os, os1, os2, os3: OctetSum;
let
  os = addOctetSum (w1.o4, w2.o4, X0);
  os1 = addOctetSum (w1.o3, w2.o3, os.x);
  os2 = addOctetSum (w1.o2, w2.o2, os1.x);
  os3 = addOctetSum (w1.o1, w2.o1, os2.x);
  ws = {x: os3.x, 
        w: {o1: os3.o, o2: os2.o, o3: os1.o, o4: os.o}};
tel;
------------------------------------------
function dropCarryBlockSum (ws: BlockSum) returns (w: Block)
  w = ws.w;
------------------------------------------
function addBlock (w1, w2 : Block) returns (w: Block)
  w = dropCarryBlockSum (addBlockSum (w1, w2));
------------------------------------------
function addBlockHalf (h1: Half; w1: Block) returns (w: Block)
  w = addBlock ({o1: x00, o2: x00, o3: h1.o1, o4: h1.o2}, w1);
------------------------------------------
function addBlockHalves (h1, h2: Half) returns (w: Block)
  w = addBlock ({o1: x00, o2: x00, o3: h1.o1, o4: h1.o2},
                {o1: x00, o2: x00, o3: h2.o1, o4: h2.o2});
------------------------------------------
\end{lstlisting}
\section{Definitions for Type Pair}
We define 64-bit words (named ``pair'' according to the MAA terminology) using a constructor buildPair that takes two blocks and returns a pair.
The main function for this type is mulBlock (which takes two blocks and computes their 64-bit product); using auxiliary functions presented in Section~\ref{sec:32bitmul} implementing a 32-bit multiplier built using four 16-bit multipliers defined in Section~\ref{sec:16bitmul}.
\begin{lstlisting}[language=Lustre]
------------------------------------------
type Pair = {w1: Block, w2: Block};
------------------------------------------
function mulBlock (w1, w2: Block) returns (ww: Pair)
var w11, w12, w21, w22, w3, w4, w5: Block;
let
  w11 = mulHalf (HalfU (w1), HalfU (w2));
  w12 = mulHalf (HalfU (w1), HalfL (w2));
  w21 = mulHalf (HalfL (w1), HalfU (w2));
  w22 = mulHalf (HalfL (w1), HalfL (w2));
  w3 = addBlockHalf (HalfL (w12),
                     addBlockHalves (HalfL (w21), HalfU (w22)));
  w4 = addBlockHalf (HalfU (w3), addBlockHalf (HalfL (w11), 
                             addBlockHalves (HalfU (w12), HalfU(w21))));
  w5 = addBlockHalves (HalfU (w4), HalfU (w11));
  ww = {w1: {o1: w5.o3, o2: w5.o4, o3: w4.o3, o4: w4.o4}, 
        w2: {o1: w3.o3, o2: w3.o4, o3: w22.o3, o4: w22.o4}};
tel;
------------------------------------------
\end{lstlisting}

\section{Definitions for Type Key}
We define a type Key that is intended to represent the 64-bit keys (J, K) used by the MAA. 
This type has a constructor buildKey that takes two blocks and returns a key. 
In~\cite{Menezes-vanOorschot-Vanstone-96}, keys are represented using the type Pair, but we prefer introducing a dedicated type to clearly distinguish between keys and, e.g., results of the multiplication of two blocks.
\begin{lstlisting}[language=Lustre]
type Key = {K: Block, J: Block};
\end{lstlisting}

\section{Definitions (1) of MAA-specific Cryptographic Functions}
\label{section:maa-related}
We define a first set of functions to be used for MAA computations, most of which were present in~\cite{Davies-Clayden-88} or have been later introduced in~\cite{Menezes-vanOorschot-Vanstone-96}. 
\begin{lstlisting}[language=Lustre]
------------------------------------------
function CYC (w1: Block) returns (w: Block)
 w = {o1: {x1: w1.o1.x2, x2: w1.o1.x3, x3: w1.o1.x4, x4: w1.o1.x5,
           x5: w1.o1.x6, x6: w1.o1.x7, x7: w1.o1.x8, x8: w1.o2.x1},
      o2: {x1: w1.o2.x2, x2: w1.o2.x3, x3: w1.o2.x4, x4: w1.o2.x5,
	       x5: w1.o2.x6, x6: w1.o2.x7, x7: w1.o2.x8, x8: w1.o3.x1},
      o3: {x1: w1.o3.x2, x2: w1.o3.x3, x3: w1.o3.x4, x4: w1.o3.x5,
	       x5: w1.o3.x6, x6: w1.o3.x7, x7: w1.o3.x8, x8: w1.o4.x1}, 
      o4: {x1: w1.o4.x2, x2: w1.o4.x3, x3: w1.o4.x4, x4: w1.o4.x5,
	       x5: w1.o4.x6, x6: w1.o4.x7, x7: w1.o4.x8, x8: w1.o1.x1}};
------------------------------------------
function FIX1 (w1: Block) returns (w: Block)
  w = andBlock (orBlock (w1, x02040801), xBFEF7FDF);
------------------------------------------
function FIX2 (w1: Block) returns (w: Block)
  w = andBlock (orBlock (w1, x00804021), x7DFEFBFF);
------------------------------------------
function needAjust (o: Octet) returns (b: bool)
  b = ((o = x00) or (o =xFF));
------------------------------------------
function adjustCode (o: Octet) returns (x: Bit)
  x = if needAjust (o) = true then X1 else X0;
------------------------------------------
function adjust (o1, o2: Octet) returns (o: Octet)
  o = if needAjust (o1) = true then xorOctet (o1, o2) else o1;
------------------------------------------
function PAT(w1, w2: Block) returns (o: Octet)
  o = {x1: adjustCode (w1.o1), x2: adjustCode (w1.o2),
       x3: adjustCode (w1.o3), x4: adjustCode (w1.o4), 
       x5: adjustCode (w2.o1), x6: adjustCode (w2.o2),
       x7: adjustCode (w2.o3), x8: adjustCode (w2.o4)};
------------------------------------------
function BYT (w1, w2: Block) returns (w, wp: Block)
var opat: Octet;
let
  opat = PAT (w1, w2);
  w = {o1: adjust (w1.o1, rightOctet7 (opat)),
       o2: adjust (w1.o2, rightOctet6 (opat)),
       o3: adjust (w1.o3, rightOctet5 (opat)),
       o4: adjust (w1.o4, rightOctet4 (opat))};
  wp = {o1: adjust (w2.o1, rightOctet3 (opat)),
        o2: adjust (w2.o2, rightOctet2 (opat)),
        o3: adjust (w2.o3, rightOctet1 (opat)),
        o4: adjust (w2.o4, opat)};
tel;
------------------------------------------
function ADDC (w1, w2: Block) returns (ww: Pair)
var ws: BlockSum;
let
  ws = addBlockSum (w1, w2);
  ww = if ws.x = X0 then {w1: x00000000, w2: ws.w} 
       else {w1: x00000001, w2: ws.w};
tel;
------------------------------------------
\end{lstlisting}

\section{Definitions (2) of MAA-specific Cryptographic Functions}
We define a second set of functions, namely the ``multiplicative'' functions used for MAA computations. 
The three principal operations are MUL1, MUL2, and MUL2A.
\begin{lstlisting}[language=Lustre]
------------------------------------------
function MUL1 (w1, w2 : Block) returns (w: Block)
var w1w2, w3w4: Pair;
let
  w1w2 = mulBlock (w1, w2);
  w3w4 = ADDC (w1w2.w1, w1w2.w2);
  w = addBlock (w3w4.w2, w3w4.w1);
tel;
------------------------------------------
function MUL2 (w1, w2 : Block) returns (w: Block)
var w1w2, w3w4, w5w6: Pair; w3: Block;
let
  w1w2 = mulBlock (w1, w2);
  w3w4 = ADDC (w1w2.w1, w1w2.w1);
  w3 = addBlock (w3w4.w2, addBlock (w3w4.w1, w3w4.w1));
  w5w6 = ADDC (w3, w1w2.w2);
  w = addBlock (w5w6.w2, addBlock (w5w6.w1, w5w6.w1));
tel;
------------------------------------------
function MUL2A (w1, w2 : Block) returns (w: Block)
var w1w2, w3w4: Pair; w3: Block;
let
  w1w2 = mulBlock (w1, w2);
  w3 = addBlock (w1w2.w1, w1w2.w1);
  w3w4 = ADDC (w3, w1w2.w2);
  w = addBlock (w3w4.w2, addBlock (w3w4.w1, w3w4.w1));
tel;
------------------------------------------
\end{lstlisting}

\section{Definitions (3) of MAA-specific Cryptographic Functions}
\label{sec:maa-mai-funct}
We define a third set of auxilary functions used for MAA computations, and the higher-level functions that implement the MAA algorithm, namely the prelude, the inner loop, and the coda; the two principal functions are MAA (which computes the signature of a nonsegmented message) and MAC (which splits a message into 1024-byte segments and computes the overall signature of this message by iterating on each segment, the 4-byte signature of each segment being prepended to the bytes of the next segment).
\begin{lstlisting}[language=Lustre]

------------------------------------------
function squareHalf (h: Half) returns (w: Block)
  w = mulHalf (h, h);
------------------------------------------
function Q (o: Octet) returns (w: Block)
  w = squareHalf (addHalf ({o1: x00, o2: o}, x0001));
------------------------------------------
function preludeJ (J1 : Block) returns (J12, J14, J16, J18: Block;
                                        J22, J24, J26, J28: Block)
let
   J12 = MUL1 (J1, J1);
   J14 = MUL1 (J12, J12);
   J16 = MUL1 (J12, J14);
   J18 = MUL1 (J12, J16);
   J22 = MUL2 (J1, J1);
   J24 = MUL2 (J22, J22);
   J26 = MUL2 (J22, J24);
   J28 = MUL2 (J22, J26);
tel;
------------------------------------------
function preludeK (K1: Block) returns (K12, K14, K15, K17, K19: Block;
                                       K22, K24, K25, K27, K29: Block)
let
   K12 = MUL1 (K1, K1);
   K14 = MUL1 (K12, K12);
   K15 = MUL1 (K1, K14);
   K17 = MUL1 (K12, K15);
   K19 = MUL1 (K12, K17);
   K22 = MUL2 (K1, K1);
   K24 = MUL2 (K22, K22);
   K25 = MUL2 (K1, K24);
   K27 = MUL2 (K22, K25);
   K29 = MUL2 (K22, K27);
tel;
------------------------------------------
function preludeHJ (J14, J16, J18, J24, J26, J28: Block)
                   returns (H4, H6, H8: Block)
let
   H4 = xorBlock (J14, J24);
   H6 = xorBlock (J16, J26);
   H8 = xorBlock (J18, J28);
tel;
------------------------------------------
function preludeHK (K15, K17, K19, K25, K27, K29: Block; P : Octet)
                    returns (H0, H5, H7, H9: Block)
let
   H0 = xorBlock (K15, K25);
   H5 = MUL2 (H0, Q (P));
   H7 = xorBlock (K17, K27);
   H9 = xorBlock (K19, K29);
tel;
------------------------------------------
function prelude (J, K: Block) returns (X, Y, V, W, S, T: Block)
var P: Octet; J1, J14, J16, J18, J24, J26, J28: Block;
    K1, K15, K17, K25, K27, K19, K29: Block;
    H4, H5, H6, H7, H8, H9: Block;
let
  J1, K1 = BYT (J, K);
  P = PAT (J, K);
  _, J14, J16, J18, _, J24, J26, J28 = preludeJ (J1);
  _, _, K15, K17, K19, _, _, K25, K27, K29 = preludeK (K1);
  H4, H6, H8 = preludeHJ (J14, J16, J18, J24, J26, J28);
  _, H5, H7, H9 = preludeHK (K15, K17, K19, K25, K27, K29, P);
  X, Y = BYT (H4, H5);
  V, W = BYT (H6, H7);
  S, T = BYT (H8, H9);
tel;
------------------------------------------
function mainLoop (X, Y, V, W, B: Block) returns (Xp, Yp, Vp: Block)
var E: Block;
let
  Vp = CYC (V);
  E = xorBlock (Vp,W);
  Xp = MUL1  (xorBlock (X, B), FIX1 (addBlock (xorBlock (Y, B), E)));
  Yp = MUL2A (xorBlock (Y, B), FIX2 (addBlock (xorBlock (X, B), E)));
tel;
------------------------------------------
function mainLoop2 (X0, Y0, V0, W, Z, B: Block) 
returns (Xp, Yp, Vp: Block)
var X, V, Y: Block;
let
  X, Y, V = mainLoop (X0, Y0, V0, W, Z);
  Xp, Yp, Vp = mainLoop (X, Y, V, W, B);
tel;
------------------------------------------
function coda (X, Y, V, W, S, T: Block) returns (Z: Block)
var X1, X2, Y1, Y2, V1: Block;
let
  X1, Y1, V1 = mainLoop (X, Y, V, W, S);
  X2, Y2, _ = mainLoop (X1, Y1, V1, W, T);
  Z = xorBlock (X2,Y2);
tel;
------------------------------------------
node MAC (KJ: Key; Mn: Block; init: bool) 
         returns (X, Y, V, W, S, T, Z: Block; n: int32)
var X0, Y0, V0: Block; 
    newSegment: bool;
let
  n = 0 -> if init then 0 else ((pre n) + 1) mod 256;
  newSegment = false -> if pre n = 255 then true else false;
  -- initialisations
  X0, Y0, V0, W, S, T = prelude (KJ.J, KJ.K);
  -- mainloops
  X, Y, V = mainLoop (X0, Y0, V0, W, Mn) -> 
              if init then 
                mainLoop (X0, Y0, V0, W, Mn)
              else if newSegment then
                -- mode of operations
                mainLoop2 (X0, Y0, V0, W, pre Z, Mn)
              else mainLoop (pre X, pre Y, pre V, W, Mn);
  -- coda
  Z = coda (X, Y, V, W, S, T);
tel;
------------------------------------------
\end{lstlisting}

\section{Test Vectors (1) for Checking MAA Computations}
We define a first set of test vectors for the MAA. The following expressions implement the checks listed in Tables 1, 2, and 3 of~\cite{Davies-Clayden-88} and should all evaluate to true if the MAA functions are correctly implemented.
\begin{lstlisting}[language=Lustre]
function CHECK_Table_1_2 () returns (res: bool)
var U, L, Up, Lp, Upp, Lpp: Block;
    t_mul1, t_mul2, t_mul2A, t_byt, t_pat: bool;
let
-- this function checks the official test vectors given 
-- in [ISO 8730:1990] on the one hand,
-- and [ISO 8731-2:1992] and [Davies-Clayden-88] on the other hand
-- test vectors for function MUL1 - cf. Table 1 of [ISO 8731-2:1992]
  t_mul1 = ((MUL1 (x0000000F, x0000000E) = x000000D2) and
            (MUL1 (xFFFFFFF0, x0000000E) = xFFFFFF2D) and
            (MUL1 (xFFFFFFF0, xFFFFFFF1) = x000000D2));
  -- test vectors for function MUL2 - cf. Table 1 of [ISO 8731-2:1992]
  t_mul2 = ((MUL2 (x0000000F, x0000000E) = x000000D2) and
            (MUL2 (xFFFFFFF0, x0000000E) = xFFFFFF3A) and
            (MUL2 (xFFFFFFF0, xFFFFFFF1) = x000000B6));
  -- test vectors for function MUL2A - cf. Table 1 of [ISO 8731-2:1992]
  t_mul2A = ((MUL2A (x0000000F, x0000000E) = x000000D2) and
             (MUL2A (xFFFFFFF0, x0000000E) = xFFFFFF3A) and
             (MUL2A (x7FFFFFF0, xFFFFFFF1) = x800000C2) and
             (MUL2A (xFFFFFFF0, x7FFFFFF1) = x000000C4));
  -- test vectors for function BYT - cf. Table 2 of [ISO 8731-2:1992]
  U, L = BYT (x00000000, x00000000);
  Up, Lp = BYT (xFFFF00FF, xFFFFFFFF);
  Upp, Lpp = BYT (xAB00FFCD, xFFEF0001);
  t_byt = (U = x0103070F) and (L = x1F3F7FFF) and (Up = xFEFC07F0) and
          (Lp = xE0C08000) and (Upp = xAB01FCCD) and (Lpp = xF2EF3501);

  -- test vectors for function PAT - cf. Table 2 of [ISO 8731-2:1992]
  t_pat = (PAT(x00000000, x00000000) = xFF) and 
          (PAT(xFFFF00FF, xFFFFFFFF) = xFF) and
          (PAT(xAB00FFCD, xFFEF0001) = x6A);
  --
  res = t_mul1 and t_mul2  and t_mul2A and t_byt and t_pat;
tel;
------------------------------------------
function CHECK_Table_3 () returns (res: bool)
var U, Up, Upp, L, Lp, Lpp: Block;
    J1, J12, J14, J16, J18, J22, J24, J26, J28: Block;
    K1, K12, K14, K15, K17, K19, K22, K24, K25, K27, K29: Block;
    H0, H4, H5, H6, H7, H8, H9: Block; P: Octet;
    t_J1i, t_J2i, t_Hi, t_K1i, t_K2i, t_Hi2, t_PAT, t_byt: bool; 
let
  J1 = x00000100;
  K1 = x00000080;
  P  = x01;
  J12, J14, J16, J18, J22, J24, J26, J28 = preludeJ (J1);
  K12, K14, K15, K17, K19, K22, K24, K25, K27, K29 = preludeK (K1);
  H4, H6, H8 = preludeHJ (J14, J16, J18, J24, J26, J28);
  H0, H5, H7, H9 = preludeHK (K15, K17, K19, K25, K27, K29, P);
  -- test vectors for J1i values - cf. Table 3 of [ISO 8731-2:1992]
  t_J1i = (J12 = x00010000) and (J14 = x00000001) and
          (J16 = x00010000) and (J18 = x00000001);
  -- test vectors for J2i values - cf. Table 3 of [ISO 8731-2:1992]
  t_J2i = (J22 = x00010000) and (J24 = x00000002) and
          (J26 = x00020000) and (J28 = x00000004);
  -- test vectors for Hi values - cf. Table 3 of [ISO 8731-2:1992]
  t_Hi = (H4 = x00000003) and (H6 = x00030000) and (H8 = x00000005);
  -- test vectors for K1i values - cf. Table 3 of [ISO 8731-2:1992]
  t_K1i = (K12 = x00004000) and (K14 = x10000000) and 
          (K15 = x00000008) and (K17 = x00020000) and
          (K19 = x80000000);
  -- test vectors for K2i values - cf. Table 3 of [ISO 8731-2:1992]
  t_K2i = (K22 = x00004000) and (K24 = x10000000) and
          (K25 = x00000010) and (K27 = x00040000) and
          (K29 = x00000002);
  -- test vectors for Hi values - cf. Table 3 of [ISO 8731-2:1992]
  t_Hi2 = (H0 = x00000018) and (Q (P) = x00000004) and
          (H5 = x00000060) and (H7 = x00060000) and
          (H9 = x80000002);
  -- test vectors for function PAT - cf. Table 3 of [ISO 8731-2:1992]
  t_PAT = (PAT (H4, H5) = xEE) and (PAT (H6, H7) = xBB)
          and (PAT (H8, H9) = xE6);
  -- test vectors for function BYT - logically inferred from Table 3
  U, L = BYT (H4, H5);
  Up, Lp = BYT (H6, H7);
  Upp, Lpp = BYT (H8, H9);
  t_byt = (U = x01030703) and (L = x1D3B7760) and (Up = x0103050B)
          and (Lp = x17065DBB) and (Upp = x01030705) and
          (Lpp = x80397302);
   --
   res = t_J1i and t_J2i and t_Hi and t_K1i and t_K2i 
         and t_K2i and t_Hi2 and t_PAT and t_byt;
tel;
\end{lstlisting}

\section{Test Vectors (2) for Checking MAA Computations}
We define a second set of test vectors for the MAA, based upon Table 4 of~\cite{Davies-Clayden-88}.
The following expressions implement six groups of checks (three single-block messages and one three-block message).
They should all evaluate to true if the main loop of MAA (as described page 10 of~\cite{Davies-Clayden-88}) is correctly implemented.
\begin{lstlisting}[language=Lustre]
function CHECK_Table_4_m1 () returns (res: bool)
var A, B, C, D, E, F, Fp, Fpp, G, Gp, Gpp, M, V, Vp: Block;
    W, X0, X, Xp, Y0, Y, Yp, Z: Block;
let
-- test vectors for function Main Loop (Table 4 of [ISO 8731-2:1992])
  -- first single-Block message
  -- input values given in Table 4
  A  = x00000004;   -- fake "A" constant
  B  = x00000001;   -- fake "B" constant
  C  = xFFFFFFF7;   -- fake "C" constant
  D  = xFFFFFFFB;   -- fake "D" constant
  V  = x00000003;
  W  = x00000003;
  X0 = x00000002;
  Y0 = x00000003;
  M  = x00000005;
  -- loop iteration described page 10 of [ISO 8731-2:1992]
  Vp = CYC (V);            
  E = xorBlock (Vp, W);    
  X = xorBlock (X0, M);    
  Y = xorBlock (Y0, M); 
  F = addBlock (E, Y); 
  G = addBlock (E, X);   
  Fp = orBlock (F, A);    
  Gp = orBlock (G, B);   
  Fpp = andBlock (Fp, C);  
  Gpp = andBlock (Gp, D); 
  Xp = MUL1 (X, Fpp);  
  Yp = MUL2A (Y, Gpp); 
  Z = xorBlock (Xp, Yp); 
  --
  res = (Vp = x00000006) and (E = x00000005) and
        (X  = x00000007) and (Y = x00000006) and
        (F  = x0000000B) and (G = x0000000C) and
        (Fp = x0000000F) and (Gp = x0000000D) and
        (Fpp = x00000007) and  (Gpp = x00000009) and
        (Xp = x00000031) and (Yp = x00000036) and
        (Z = x00000007);
tel;
------------------------------------------
function CHECK_Table_4_m2 () returns (res: bool)
var A, B, C, D, E, F, Fp, Fpp, G, Gp, Gpp, M, V, Vp: Block;
    W, X0, X, Xp, Y0, Y, Yp, Z: Block;
let
-- test vectors for function Main Loop (Table 4 of [ISO 8731-2:1992])
  -- second single-Block message
  -- input values given in Table 4
  A  = x00000001;   -- fake "A" constant
  B  = x00000004;   -- fake "B" constant
  C  = xFFFFFFF9;   -- fake "C" constant
  D  = xFFFFFFFC;   -- fake "D" constant
  V  = x00000003;
  W  = x00000003;
  X0 = xFFFFFFFD;
  Y0 = xFFFFFFFC;
  M  = x00000001;
  -- loop iteration described page 10 of [ISO 8731-2:1992]
  Vp = CYC (V);           
  E = xorBlock (Vp, W);   
  X = xorBlock (X0, M);  
  Y = xorBlock (Y0, M);   
  F = addBlock (E, Y);    
  G = addBlock (E, X);  
  Fp = orBlock (F, A);    
  Gp = orBlock (G, B);     
  Fpp = andBlock (Fp, C);
  Gpp = andBlock (Gp, D); 
  Xp = MUL1 (X, Fpp);  
  Yp = MUL2A (Y, Gpp);  
  Z = xorBlock (Xp, Yp);
  --
  res = (Vp = x00000006) and (E = x00000005) and (X = xFFFFFFFC)
        and (Y = xFFFFFFFD) and (F = x00000002) and (G = x00000001)
        and (Fp = x00000003) and (Gp = x00000005) and (Fpp = x00000001)
        and (Gpp = x00000004) and (Xp = xFFFFFFFC) and (Yp = xFFFFFFFA)
        and (Z = x00000006);
tel;
------------------------------------------
function CHECK_Table_4_m3 () returns (res: bool)
var A, B, C, D, E, F, Fp, Fpp, G, Gp, Gpp, M, V, Vp: Block;
    W, X0, X, Xp, Y0, Y, Yp, Z: Block;
let
-- test vectors for function Main Loop (Table 4 of [ISO 8731-2:1992])
  -- third single-Block message
  -- input values given in Table 4
  A  = x00000001;   -- fake "A" constant
  B  = x00000002;   -- fake "B" constant
  C  = xFFFFFFFE;   -- fake "C" constant
  D  = x7FFFFFFD;   -- fake "D" constant
  V  = x00000007;
  W  = x00000007;
  X0 = xFFFFFFFD;
  Y0 = xFFFFFFFC;
  M  = x00000008;
  -- loop iteration described page 10 of [ISO 8731-2:1992]
  Vp = CYC (V);          
  E = xorBlock (Vp, W);   
  X = xorBlock (X0, M);   
  Y = xorBlock (Y0, M);  
  F = addBlock (E, Y);    
  G = addBlock (E, X);   
  Fp = orBlock (F, A);    
  Gp = orBlock (G, B);    
  Fpp = andBlock (Fp, C); 
  Gpp = andBlock (Gp, D);  
  Xp = MUL1 (X, Fpp);      
  Yp = MUL2A (Y, Gpp);   
  Z = xorBlock (Xp, Yp);
  --
  res = (Vp = x0000000E) and (E = x00000009) and (X = xFFFFFFF5)
        and (Y = xFFFFFFF4) and (F = xFFFFFFFD) and (G = xFFFFFFFE)
        and (Fp = xFFFFFFFD) and (Gp = xFFFFFFFE) and (Fpp = xFFFFFFFC)
        and (Gpp = x7FFFFFFC) and (Xp = x0000001E) and (Yp = x0000001E)
        and (Z = x00000000);
tel;
------------------------------------------
function CHECK_3_messages_m1 () returns (res: bool)
var A, B, C, D, E, F, Fp, Fpp, G, Gp, Gpp, M, V, Vp: Block;
    W, X0, X, Xp, Y0, Y, Yp, Z: Block;
let
-- three-Block message: first Block
  -- input values given in Table 4
  A  = x00000002;   -- fake "A" constant
  B  = x00000001;   -- fake "B" constant
  C  = xFFFFFFFB;   -- fake "C" constant
  D  = xFFFFFFFB;   -- fake "D" constant
  V  = x00000001;
  W  = x00000001;
  X0 = x00000001;
  Y0 = x00000002;
  M  = x00000000;
  -- loop iteration described page 10 of [ISO 8731-2:1992]
  Vp = CYC (V);   
  E = xorBlock (Vp, W);   
  X = xorBlock (X0, M); 
  Y = xorBlock (Y0, M);  
  F = addBlock (E, Y);   
  G = addBlock (E, X);   
  Fp = orBlock (F, A);  
  Gp = orBlock (G, B);   
  Fpp = andBlock (Fp, C);
  Gpp = andBlock (Gp, D); 
  Xp = MUL1 (X, Fpp);    
  Yp = MUL2A (Y, Gpp);    
  Z = xorBlock (Xp, Yp);  
  --
  res = (Vp = x00000002) and (E = x00000003) and (X = x00000001)
        and (Y = x00000002) and (F = x00000005) and (G = x00000004)
        and (Fp = x00000007) and (Gp = x00000005) and (Fpp = x00000003)
        and (Gpp = x00000001) and (Xp = x00000003) and (Yp = x00000002)
        and (Z = x00000001);
tel;
------------------------------------------
function CHECK_3_messages_m2 () returns (res: bool)
var A, B, C, D, E, F, Fp, Fpp, G, Gp, Gpp, M, V, Vp: Block;
    W, X0, X, Xp, Y0, Y, Yp, Z: Block;
let
-- three-Block message: second Block
  -- input values given in Table 4
  A  = x00000002;   -- fake "A" constant
  B  = x00000001;   -- fake "B" constant
  C  = xFFFFFFFB;   -- fake "C" constant
  D  = xFFFFFFFB;   -- fake "D" constant
  V  = x00000002;
  W  = x00000001;
  X0 = x00000003;
  Y0 = x00000002;
  M  = x00000001;
  -- loop iteration described page 10 of [ISO 8731-2:1992]
  Vp = CYC (V);   
  E = xorBlock (Vp, W);  
  X = xorBlock (X0, M);   
  Y = xorBlock (Y0, M);  
  F = addBlock (E, Y);  
  G = addBlock (E, X);   
  Fp = orBlock (F, A);    
  Gp = orBlock (G, B);   
  Fpp = andBlock (Fp, C);  
  Gpp = andBlock (Gp, D); 
  Xp = MUL1 (X, Fpp);     
  Yp = MUL2A (Y, Gpp);     
  Z = xorBlock (Xp, Yp);  
  --
  res = (Vp = x00000004) and (E = x00000005) and (X = x00000002)
        and (Y = x00000003) and (F = x00000008) and (G = x00000007)
        and (Fp = x0000000A) and (Gp = x00000007) and (Fpp = x0000000A)
        and (Gpp = x00000003) and (Xp = x00000014) and (Yp = x00000009)
        and (Z = x0000001D);
tel;
------------------------------------------
function CHECK_3_messages_m3 () returns (res: bool)
var A, B, C, D, E, F, Fp, Fpp, G, Gp, Gpp, M, V, Vp: Block;
    W, X0, X, Xp, Y0, Y, Yp, Z: Block;
let
-- three-Block message: third Block
  -- input values given in Table 4
  A  = x00000002;   -- fake "A" constant
  B  = x00000001;   -- fake "B" constant
  C  = xFFFFFFFB;   -- fake "C" constant
  D  = xFFFFFFFB;   -- fake "D" constant
  V  = x00000004;
  W  = x00000001;
  X0 = x00000014;
  Y0 = x00000009;
  M  = x00000002;
  -- loop iteration described page 10 of [ISO 8731-2:1992]
  Vp = CYC (V);         
  E = xorBlock (Vp, W);  
  X = xorBlock (X0, M);   
  Y = xorBlock (Y0, M);
  F = addBlock (E, Y); 
  G = addBlock (E, X);   
  Fp = orBlock (F, A);    
  Gp = orBlock (G, B);    
  Fpp = andBlock (Fp, C); 
  Gpp = andBlock (Gp, D); 
  Xp = MUL1 (X, Fpp);    
  Yp = MUL2A (Y, Gpp);   
  Z = xorBlock (Xp, Yp);  
  --
  res = (Vp = x00000008) and (E = x00000009) and (X = x00000016)
        and (Y = x0000000B) and (F = x00000014) and (G = x0000001F)
        and (Fp = x00000016) and (Gp = x0000001F) and (Fpp = x00000012)
        and (Gpp = x0000001B) and (Xp = x0000018C) and (Yp = x00000129)
        and (Z = x000000A5);
tel;
------------------------------------------
function CHECK_Annex_E () returns (res: bool)
var A, B, C, D, E, F, Fp, Fpp, G, Gp, Gpp, M, V0, V: Block;
    W, X0, X, Xp, Y0, Y, Yp: Block;
let
-- test vectors of Annex E.3.3 of [ISO 8730:1990]
  A  = x02040801;   -- true "A" constant
  B  = x00804021;   -- true "B" constant
  C  = xBFEF7FDF;   -- true "C" constant
  D  = x7DFEFBFF;   -- true "D" constant
  X0 = x21D869BA;
  Y0 = x7792F9D4;
  V0 = xC4EB1AEB;
  W  = xF6A09667;
  M  = x0A202020;
  -- loop iteration on the first Block M
  V = CYC (V0);     
  E = xorBlock (V, W);    
  X = xorBlock (X0, M);  
  Y = xorBlock (Y0, M); 
  F = addBlock (E, Y);  
  G = addBlock (E, X); 
  Fp = orBlock (F, A); 
  Gp = orBlock (G, B);  
  Fpp = andBlock (Fp, C);
  Gpp = andBlock (Gp, D);  
  Xp = MUL1 (X, Fpp);   
  Yp = MUL2A (Y, Gpp); 
  -- 
  res = (V = x89D635D7) and (E = x7F76A3B0) and (X = x2BF8499A)
        and (Y = x7DB2D9F4) and (F = xFD297DA4) and (G = xAB6EED4A)
        and (Fp = xFF2D7DA5) and (Gp = xABEEED6B) and (Fpp = xBF2D7D85)
        and (Gpp = x29EEE96B) and (Xp = x0AD67E20) and (Yp = x30261492);
tel;
------------------------------------------
\end{lstlisting}
We complete the above tests with additional test vectors taken from~\cite[Annex~E.3.3]{ISO-8730:1990}, which only gives detailed values for the first block of the 84-block test message.
\begin{lstlisting}[language=Lustre]
function CHECK_Annex_E () returns (res: bool)
var A, B, C, D, E, F, Fp, Fpp, G, Gp, Gpp, M, V0, V: Block;
    W, X0, X, Xp, Y0, Y, Yp: Block;
let
-- test vectors of Annex E.3.3 of [ISO 8730:1990]
  A  = x02040801;   -- true "A" constant
  B  = x00804021;   -- true "B" constant
  C  = xBFEF7FDF;   -- true "C" constant
  D  = x7DFEFBFF;   -- true "D" constant
  X0 = x21D869BA;
  Y0 = x7792F9D4;
  V0 = xC4EB1AEB;
  W  = xF6A09667;
  M  = x0A202020;
  -- loop iteration on the first Block M
  V = CYC (V0);     
  E = xorBlock (V, W);    
  X = xorBlock (X0, M);  
  Y = xorBlock (Y0, M); 
  F = addBlock (E, Y);  
  G = addBlock (E, X); 
  Fp = orBlock (F, A); 
  Gp = orBlock (G, B);  
  Fpp = andBlock (Fp, C);
  Gpp = andBlock (Gp, D);  
  Xp = MUL1 (X, Fpp);   
  Yp = MUL2A (Y, Gpp); 
  -- 
  res = (V = x89D635D7) and (E = x7F76A3B0) and (X = x2BF8499A)
        and (Y = x7DB2D9F4) and (F = xFD297DA4) and (G = xAB6EED4A)
        and (Fp = xFF2D7DA5) and (Gp = xABEEED6B) and (Fpp = xBF2D7D85)
        and (Gpp = x29EEE96B) and (Xp = x0AD67E20) and (Yp = x30261492);
tel;
------------------------------------------
\end{lstlisting}

\section{Test vectors (3) for Checking MAA Computations}
We define a third set of test vectors for the MAA, based upon Table 5 of~\cite{Davies-Clayden-88}.
The following expressions implement four groups of checks, with two different keys and two different messages. 
They should all evaluate to true if the MAA signature is correctly computed.
\begin{lstlisting}[language=Lustre]
function CHECK_Table_5_v1 () returns (res: bool)
var J, K, X0, X, Xp, Xpp, Xppp, Y0, Y, Yp, Ypp, Yppp: Block;
    V0, V, Vp, Vpp, W, S, T, Z, M1, M2: Block;
let
-- test vectors for the whole algorithm (Table 5 of [ISO 8731-2:1992])
  -- first column of Table 5
  J  = x00FF00FF;
  K  = x00000000;
  M1 = x55555555;
  M2 = xAAAAAAAA;
  X0, Y0, V0, W, S, T = prelude (J, K);
  -- 1st MainLoop iteration
  X, Y, V = mainLoop (X0, Y0, V0, W, M1);
  -- 2nd MainLoop iteration
  Xp, Yp, Vp = mainLoop (X, Y, V, W, M2);
  -- Coda: MainLoop iteration with S
  Xpp, Ypp, Vpp = mainLoop (Xp, Yp, Vp, W, S);
  -- Coda: MainLoop iteration with T
  Xppp, Yppp, _ = mainLoop (Xpp, Ypp, Vpp, W, T);
  Z  = xorBlock (Xppp,Yppp);
  --
  res = (PAT (J, K) = xFF) and (X0 = x4A645A01) and (Y0 = x50DEC930) and
        (V0 = x5CCA3239) and (W = xFECCAA6E) and (S = x51EDE9C7) and
        (T = x24B66FB5) and (X = x48B204D6) and (Y = x5834A585) and
        (Xp = x4F998E01) and (Yp = xBE9F0917) and (Ypp = xDB9102B0) and 
        (Xpp = x344925FC) and (Xppp = x277B4B25) and (Yppp = xD636250D)
        and (Z = xF14D6E28);
tel;
------------------------------------------
function CHECK_Table_5_v2 () returns (res: bool)
var J, K, X0, X, Xp, Xpp, Xppp, Y0, Y, Yp, Ypp, Yppp: Block;
    V0, V, Vp, Vpp, W, S, T, Z, M1, M2: Block;
let
-- test vectors for the whole algorithm (Table 5 of [ISO 8731-2:1992])
  -- second column of Table 5
  J  = x00FF00FF;
  K  = x00000000;
  M1 = xAAAAAAAA;
  M2 = x55555555;
  X0, Y0, V0, W, S, T = prelude (J, K);
  -- 1st MainLoop iteration
  X, Y, V = mainLoop (X0, Y0, V0, W, M1);
  -- 2nd MainLoop iteration
  Xp, Yp, Vp = mainLoop (X, Y, V, W, M2);
  -- Coda: MainLoop iteration with S
  Xpp, Ypp, Vpp = mainLoop (Xp, Yp, Vp, W, S);
  -- Coda: MainLoop iteration with T
  Xppp, Yppp, _ = mainLoop (Xpp, Ypp, Vpp, W, T);
  Z  = xorBlock (Xppp,Yppp);
  --
  res = (PAT (J, K) = xFF) and (X0 = x4A645A01) and (Y0 = x50DEC930)
        and (V0 = x5CCA3239) and  (W = xFECCAA6E) and (S = x51EDE9C7)
        and (T = x24B66FB5) and (X = x6AEBACF8) and (Y = x9DB15CF6)
        and (Xp = x270EEDAF) and (Yp = xB8142629) and (Xpp = x29907CD8)
        and (Ypp = xBA92DB12) and (Xppp = x28EAD8B3) and 
        (Yppp = x81D10CA3) and (Z = xA93BD410);
tel;
------------------------------------------
function CHECK_Table_5_v3 () returns (res: bool)
var J, K, X0, X, Xp, Xpp, Xppp, Y0, Y, Yp, Ypp, Yppp: Block;
    V0, V, Vp, Vpp, W, S, T, Z, M1, M2: Block;
let
-- test vectors for the whole algorithm (Table 5 of [ISO 8731-2:1992])
  -- third column of Table 5
  J  = x55555555;
  K  = x5A35D667;
  M1 = x00000000;
  M2 = xFFFFFFFF;
  X0, Y0, V0, W, S, T = prelude (J, K);
  -- 1st MainLoop iteration
  X, Y, V = mainLoop (X0, Y0, V0, W, M1);
  -- 2nd MainLoop iteration
  Xp, Yp, Vp = mainLoop (X, Y, V, W, M2);
  -- Coda: MainLoop iteration with S
  Xpp, Ypp, Vpp = mainLoop (Xp, Yp, Vp, W, S);
  -- Coda: MainLoop iteration with T
  Xppp, Yppp, _ = mainLoop (Xpp, Ypp, Vpp, W, T);
  Z  = xorBlock (Xppp,Yppp);
  --
  res = (PAT (J, K) = x00) and (X0 = x34ACF886) and (Y0 = x7397C9AE) and
        (V0 = x7201F4DC) and (W = x2829040B) and (S = x9E2E7B36) and 
        (T = x13647149) and (X = x2FD76FFB) and (Y = x550D91CE) and
        (Xp = xA70FC148) and (Yp = x1D10D8D3) and (Xpp = xB1CC1CC5)
        and (Ypp = x29C1485F) and (Xppp = x288FC786) and 
        (Yppp = x9115A558) and (Z = xB99A62DE);
tel;
------------------------------------------
function CHECK_Table_5_v4 () returns (res: bool)
var J, K, X0, X, Xp, Xpp, Xppp, Y0, Y, Yp, Ypp, Yppp: Block;
    V0, V, Vp, Vpp, W, S, T, Z, M1, M2: Block;
let
-- test vectors for the whole algorithm  (Table 5 of [ISO 8731-2:1992])
  -- fourth column of Table 5
  J = x55555555;
  K = x5A35D667;
  M1 = xFFFFFFFF;
  M2 = x00000000;
  X0, Y0, V0, W, S, T = prelude (J, K);
  -- 1st MainLoop iteration
  X, Y, V = mainLoop (X0, Y0, V0, W, M1);
  -- 2nd MainLoop iteration
  Xp, Yp, Vp = mainLoop (X, Y, V, W, M2);
  -- Coda: MainLoop iteration with S
  Xpp, Ypp, Vpp = mainLoop (Xp, Yp, Vp, W, S);
  Xppp, Yppp, _ = mainLoop (Xpp, Ypp, Vpp, W, T);
  Z = xorBlock (Xppp,Yppp);
  --
  res = (PAT (J, K) = x00) and (X0 = x34ACF886) and (Y0 = x7397C9AE) and 
        (V0 = x7201F4DC) and (W = x2829040B) and (S = x9E2E7B36) and 
        (T = x13647149) and (X = x8DC8BBDE) and (Y = xFE4E5BDD) and
        (Xp = xCBC865BA) and (Yp = x0297AF6F) and (Xpp = x3CF3A7D2) and
        (Ypp = x160EE9B5) and (Xppp = xD0482465) and (Yppp = x7050EC5E)
        and (Z = xA018C83B);
tel;
------------------------------------------
\end{lstlisting}
We complete the above tests with additional test vectors taken from from~\cite[Annex~E.3.3]{ISO-8730:1990}, which gives prelude results computed for another key.
\begin{lstlisting}[language=Lustre]
function CHECK_PRELUDE_Annex_E33 () returns (res: bool)
var J, K, X, Y, V, W, S, T: Block;
let
-- test vectors of Annex E.3.3 of [ISO 8730:1990]
  J  = xE6A12F07;
  K  = x9D15C437;
  X, Y, V, W, S, T = prelude (J, K);
  --
  res = (X = x21D869BA) and (Y = x7792F9D4) and (V = xC4EB1AEB) and
        (W = xF6A09667) and (S = x6D67E884) and (T = xA511987A);
tel;
------------------------------------------
\end{lstlisting}

\section{Test Vectors (4) for Checking MAA Computations}
\label{ap:20words}
We define a last set of test vectors for the MAA. The first one (a message of 20 blocks containing only zeros) was directly taken from Table 6 of~\cite{Davies-Clayden-88}.
\begin{lstlisting}[language=Lustre]
function CHECK_ALL_ALGO_20m () returns (res: bool)
var B, J, K, X0, Y0, V0, W, S, T: Block; tp: bool;
    X, X1, X2, X3, X4, X5, X6, X7, X8, X9, X10, X11: Block;
    X12, X13, X14, X15, X16, X17, X18, X19, X20, X21: Block;
    Y, Y1, Y2, Y3, Y4, Y5, Y6, Y7, Y8, Y9, Y10, Y11: Block;
    Y12, Y13, Y14, Y15, Y16, Y17, Y18, Y19, Y20, Y21: Block;
    V, V1, V2, V3, V4, V5, V6, V7, V8, V9, V10, V11: Block;
    V12, V13, V14, V15, V16, V17, V18, V19, V20: Block;
let
  -- test vectors for the whole algorithm
  J = x80018001;
  K = x80018000;
  -- test mentioned in Table 6 of [ISO 8731-2:1992]
  -- iterations on a message containg 20 null Blocks
  X0, Y0, V0, W, S, T = prelude (J, K);
  B = x00000000;
  -- 1st MainLoop iteration
  X, Y, V = mainLoop (X0, Y0, V0, W, B);
  -- 2nd MainLoop iteration
  X1, Y1, V1 = mainLoop (X, Y, V, W, B);
  -- 3rd MainLoop iteration
  X2, Y2, V2 = mainLoop (X1, Y1, V1, W, B);
  -- 4th MainLoop iteration
  X3, Y3, V3 = mainLoop (X2, Y2, V2, W, B);
  -- 5th MainLoop iteration
  X4, Y4, V4 = mainLoop (X3, Y3, V3, W, B);
  -- 6th MainLoop iteration
  X5, Y5, V5 = mainLoop (X4, Y4, V4, W, B);
  -- 7th MainLoop iteration
  X6, Y6, V6 = mainLoop (X5, Y5, V5, W, B);
  -- 8th MainLoop iteration
  X7, Y7, V7 = mainLoop (X6, Y6, V6, W, B);
  -- 9th MainLoop iteration
  X8, Y8, V8 = mainLoop (X7, Y7, V7, W, B);
  -- 10th MainLoop iteration
  X9, Y9, V9 = mainLoop (X8, Y8, V8, W, B);
  -- 11th MainLoop iteration
  X10, Y10, V10 = mainLoop (X9, Y9, V9, W, B);
  -- 12th MainLoop iteration
  X11, Y11, V11 = mainLoop (X10, Y10, V10, W, B);
  -- 13th MainLoop iteration
  X12, Y12, V12 = mainLoop (X11, Y11, V11, W, B);
  -- 14th MainLoop iteration
  X13, Y13, V13 = mainLoop (X12, Y12, V12, W, B);
  -- 15th MainLoop iteration
  X14, Y14, V14 = mainLoop (X13, Y13, V13, W, B);
  -- 16th MainLoop iteration
  X15, Y15, V15 = mainLoop (X14, Y14, V14, W, B);
  -- 17th MainLoop iteration
  X16, Y16, V16 = mainLoop (X15, Y15, V15, W, B);
  -- 18th MainLoop iteration
  X17, Y17, V17 = mainLoop (X16, Y16, V16, W, B);
  -- 19th MainLoop iteration
  X18, Y18, V18 = mainLoop (X17, Y17, V17, W, B);
  -- 20th MainLoop iteration
  X19, Y19, V19 = mainLoop (X18, Y18, V18, W, B);
  -- Coda: MainLoop iteration with S
  X20, Y20, V20 = mainLoop (X19, Y19, V19, W, S);
  -- Coda: MainLoop iteration with T
  X21, Y21, _ = mainLoop (X20, Y20, V20, W, T);
  --
  tp = (X = x303FF4AA) and (Y = x1277A6D4) and (X1 = x55DD063F) and
       (Y1 = x4C49AAE0) and (X2 = x51AF3C1D) and (Y2 = x5BC02502) and
       (X3 = xA44AAAC0) and (Y3 = x63C70DBA) and (X4 = x4D53901A) and
       (Y4 = x2E80AC30) and (X5 = x5F38EEF1) and (Y5 = x2A6091AE) and
       (X6 = xF0239DD5) and (Y6 = x3DD81AC6) and (X7 = xEB35B97F) and
       (Y7 = x9372CDC6) and (X8 = x4DA124A1) and (Y8 = xC6B1317E) and
       (X9 = x7F839576) and (Y9 = x74B39176) and (X10 = x11A9D254) and
       (Y10 = xD78634BC) and (X11 = xD8804CA5) and (Y11 = xFDC1A8BA) and
       (X12 = x3F6F7248) and (Y12 = x11AC46B8) and (X13 = xACBC13DD) and
       (Y13 = x33D5A466) and (X14 = x4CE933E1) and (Y14 = xC21A1846) and
       (X15 = xC1ED90DD) and (Y15 = xCD959B46) and (X16 = x3CD54DEB) and
       (Y16 = x613F8E2A) and (X17 = xBBA57835) and (Y17 = x07C72EAA) and
       (X18 = xD7843FDC) and (Y18 = x6AD6E8A4) and (X19 = x5EBA06C2) and
       (Y19 = x91896CFA) and (X20 = x1D9C9655) and (Y20 = x98D1CC75) and
       (X21 = x7BC180AB) and (Y21 = xA0B87B77);
  res = tp and (coda (X19, Y19, V19, W, S, T) = xDB79FBDC);
tel;
\end{lstlisting}